\documentclass[acmlarge]{acmart}

\AtBeginDocument{%
  \providecommand\BibTeX{{%
    \normalfont B\kern-0.5em{\scshape i\kern-0.25em b}\kern-0.8em\TeX}}}

\usepackage{graphicx}
\usepackage{subfig}
\usepackage{float}

\usepackage[normalem]{ulem}

\usepackage{longtable}

\usepackage{array, makecell, multirow, booktabs}
\newcolumntype{P}[1]{>{\centering\arraybackslash}p{#1}}
\setcellgapes{3pt}

\usepackage{pifont}
\newcommand{\cmark}{\ding{51}}
\newcommand{\xmark}{\ding{55}}

\setcopyright{none}

\usepackage{atbegshi,picture}
\usepackage{lipsum}

\AtBeginShipout{\AtBeginShipoutUpperLeft{%
  \put(\dimexpr\paperwidth-1cm\relax,-1.5cm){\makebox[0pt][r]{\framebox{This paper has been accepted for publication by the ACM Computing Surveys. The final version will be published by the ACM.}}}%
}}

\begin{document}

\title{Access Control Mechanisms in Named Data Networks: A Comprehensive Survey}

\author{Boubakr Nour}
\email{n.boubakr@ieee.org}
\orcid{1234-5678-9012}
\affiliation{
	\institution{Beijing Institute of Technology}
	\city{Beijing}
	\country{China}	
}

\author{Hakima Khelifi}
\email{hakima@bit.edu.cn}
\affiliation{
	\institution{Beijing Institute of Technology}
	\city{Beijing}
	\country{China}
}

\author{Rasheed Hussain}
\email{r.hussain@innopolis.ru}
\affiliation{
	\institution{Innopolis University}
	\city{Innopolis}
	\country{Russia}
}

\author{Spyridon Mastorakis}
\email{smastorakis@unomaha.edu}
\affiliation{
	\institution{University of Nebraska Omaha}
	\city{Omaha}
	\country{USA}
}

\author{Hassine Moungla}
\email{hassine.moungla@parisdescartes.fr}
\affiliation{
	\institution{Universit\'e de Paris}
	\city{Paris}
	\country{France}
}

\renewcommand{\shortauthors}{B. Nour et al. Survey on Named AC}

\begin{abstract}
	Information-Centric Networking (ICN) has recently emerged as a prominent candidate for the Future Internet Architecture (FIA) that addresses existing issues with the host-centric communication model of the current TCP/IP-based Internet. Named Data Networking (NDN) is one of the most recent and active ICN architectures that provides a clean slate approach for Internet communication. NDN provides intrinsic content security where security is directly provided to the content instead of communication channel. Among other security aspects, Access Control (AC) rules specify the privileges for the entities that can access the content. In TCP/IP-based AC systems, due to the client-server communication model, the servers control which client can access a particular content. In contrast, ICN-based networks use content names to drive communication and decouple the content from its original location. This phenomenon leads to the loss of control over the content causing different challenges for the realization of efficient AC mechanisms. To date, considerable efforts have been made to develop various AC mechanisms in NDN. In this paper, we provide a detailed and comprehensive survey of the AC mechanisms in NDN. We follow a holistic approach towards AC in NDN where we first summarize the ICN paradigm, describe the changes from channel-based security to content-based security and highlight different cryptographic algorithms and security protocols in NDN. We then classify the existing AC mechanisms into two main categories: \textit{Encryption-based AC} and \textit{Encryption-independent AC}. Each category has different classes based on the working principle of AC (\textit{e.g.,} Attribute-based AC, Name-based AC, Identity-based AC, etc). Finally, we present the lessons learned from the existing AC mechanisms and identify the challenges of NDN-based AC at large, highlighting future research directions for the community.
\end{abstract}



\maketitle

\section{Introduction}
\label{sec:introduction}
The current Internet ecosystem has been designed to achieve end-to-end communication between two known devices. Accordingly, each device is assigned an unique Internet Protocol (IP) address to allow communication and resources sharing. The current Internet follows the host-centric model, where the communication is based upon, among other factors, the address of the destination node. This model has been adopted as the main communication paradigm over years, where resource sharing between two devices was required. However, today's Internet is witnessing a tremendous growth in connected devices and major changes in the application design~\cite{shi2016edge}, which affect the end-user requirements. To address these changes, the IP stack has been refined with add-ons and protocols to support various features. However, TCP/IP became a complex network architecture by adding extra patches to support security, mobility, and management.

To meet the user demands and address the changes in the nature of applications, various solutions have been proposed to realize Future Internet Architectures (FIA)~\cite{pan2011survey}. The current Internet model is shifting from the host-based communication toward the content-oriented model~\cite{choi2011survey}. End-users are inherently interested in what they request and consume regardless of who is offering the content or service. This vision of communication is known as Information-Centric Networking (ICN)~\cite{xylomenos2014survey, fang2015survey, jiang2015survey}.

In contrast to the host-centric model, where an IP address is used to deliver packets to the destination host, ICN uses the content name to drive the communication and fetch the content from the network~\cite{bari2012survey}. Various ICN projects have been proposed as part of FIA including Data Oriented Network (DONA)~\cite{koponen2007data}, Scalable and Adaptive Internet Solutions (4WARD/SAIL)~\cite{4ward, sail}, Publish-Subscribe Internet Routing Paradigm (PSIRP/PURSUIT)~\cite{fotiou2010developing}, Content-Centric Networking (CCN)~\cite{oehlmann2013content}, COMET~\cite{comet}, CONVERGENCE~\cite{convergence}, MobilityFirst~\cite{mobilityfirst}, and Named-Data Networking (NDN)~\cite{zhang2014named}. We refer interested readers to other surveys~\cite{ahlgren2012survey, xylomenos2014survey} for more details. In this work, we focus on research relevant to NDN, since it has received considerable attention from the research community among various FIA proposals and it continues to be favored as the most prominent FIA candidate.

In a nutshell, NDN~\cite{zhang2010named} identifies the content using semantically meaningful, hierarchical names. NDN decouples the content from its original location and enables in-network caching. Intermediate nodes have the ability to cache content and fulfill future demands~\cite{seetharam2018caching, ullah2020icn}. NDN offers a receiver-driven communication model based on the exchange of interest-Data packets. A content consumer initiates a request in the form of an {\it interest} packet that carries the requested content name. Intermediate nodes forward the requests based on that name using name-based forwarding rules. When the request reaches a content provider/producer or a content store that can offer the requested content, a reply is sent out in the form of a {\it Data} packet.

NDN features a content-based security model~\cite{ghali2019content, zhang2018overview, yu2018content, wang2018securing}, since each Data packet is cryptographically signed by the entity that produced it. In contrast to host-based networks, where communication security relies on the security of the communication channel itself, NDN directly secures the content exchanged over the network regardless of the used communication channel. All security-related information is bound to the content and stays with the content during transmission over the network and at rest (e.g., when cached in the content store of intermediate nodes). Although this concept is promising in terms of securing content at the publishing phase, various issues may arise regarding efficiency and scalability, the used security algorithms, and privacy~\cite{ngai2017can}.

Access Control (AC)~\cite{van2014encyclopedia} is a vital aspect of today's Internet, since it determines who can access what content. Existing TCP/IP-based AC systems are heavily influenced by the client-server communication model. The request is sent to the service/content provider or a delegated server who has a list of access rules--based on this list, the privileged access is determined. This AC model contradicts the NDN communication model, where content can be cached in the network and can be retrieved from any entity that can provide it, being fundamentally different than the existing client-server communication model. 


Various efforts have been conducted by the research community to explore AC mechanisms for NDN. Some solutions take the communication back to a client-server model by enforcing the consumer to communicate with the original producer to get authenticated through access rules. However, this concept ignores the advantages of ubiquitous in-network caching and may fail due to producer's unavailability. Other solutions take advantage of NDN naming, introducing Name-based Access Control (NAC)~\cite{yu2015name}. Between the former and the latter solutions, hybrid solutions have also been proposed.

Overall, AC solutions offer the means for secure content dissemination and privileged access in NDN. Motivated by the significance of access control as a problem, in this work, we provide a comprehensive survey on the existing AC schemes in NDN.

\begin{table}[!t]
	\makegapedcells
	\centering
	\caption{Comparison of contributions among related surveys.}
	\label{tab:related_surveys}
	\scriptsize
	\begin{tabular}{
			c
			c
			P{0.09\textwidth}
			P{0.065\textwidth}
			P{0.1\textwidth}
			P{0.068\textwidth}
			p{0.3\textwidth}
			P{0.065\textwidth}
			P{0.06\textwidth}}
		\toprule
		\textbf{Ref.} &
		\textbf{Year} &
		\textbf{AC Mechanisms} &
		\textbf{Recent Work}&
		\textbf{Research Challenges} &
		\textbf{Future Direction}&
		\makecell[c]{\textbf{Limitation}}&
		\textbf{Covered AC Sol.}&
		\textbf{Covered Year}
		\\ \midrule
		
		\cite{abdallah2015survey}&
		2015 &
		\cmark &
		\xmark &
		\cmark &
		\xmark &
		$\bullet$ Broader security attack analysis. \newline
		$\bullet$ Targeting several ICN architectures. \newline
		$\bullet$ Not specialized on Named Data Networking. \newline &
		1 &
		2012 \\
		
		\cite{ambrosin2018security}&
		2018 &
		\xmark &
		\xmark &
		\cmark &
		\cmark &
		$\bullet$ No review of technical solutions. \newline
		$\bullet$ Broader security and privacy analysis. \newline
		$\bullet$ Targeting several ICN architectures. \newline
		$\bullet$ Not specialized on Named Data Networking. \newline
		$\bullet$ Not specialized on access control. &
		NA &
		NA \\
		
		\cite{tourani2018security}&
		2018 &
		\cmark &
		\xmark &
		\xmark &
		\cmark &
		$\bullet$ Focus on security, privacy, and access control. \newline
		$\bullet$ Targeting several ICN architectures. \newline
		$\bullet$ Not specialized on Named Data Networking. &
		19 &
		2009-2017 \\
		
		\cite{lutz2016security} &
		2016 &
		\xmark &
		\xmark &
		\cmark &
		\cmark &
		$\bullet$ No review of technical solutions. \newline
		$\bullet$ Not specialized on access control. &
		NA &
		NA\\
		
		\cite{ghali2017encryption}&
		2017 &
		\xmark &
		\xmark &
		\cmark &
		\cmark &
		$\bullet$ Focus on privacy in CCN networks. \newline
		$\bullet$ Not specialized on access control. &
		NA &
		NA\\
		
		\cite{bouk2018named}&
		2018 &
		\xmark &
		\xmark &
		\cmark &
		\cmark &
		$\bullet$ Focus on security attacks in vehicular cyber-physical systems. \newline
		$\bullet$ No review of technical solutions. \newline
		$\bullet$ Not specialized on access control. &
		NA &
		NA \\
		
		\cite{khelifi2018security}&
		2018 &
		\xmark &
		\xmark &
		\cmark &
		\cmark &
		$\bullet$ Focus on security and privacy in vehicular named networks. \newline
		$\bullet$ Not specialized on access control. &
		NA &
		NA \\
		
		Our &
		2019 &
		\cmark &
		\cmark &
		\cmark &
		\cmark &
		\makecell[c]{/} &
		28 &
		2009-2019 \\
		
		\bottomrule
	\end{tabular}
\end{table}

\subsection{Related Surveys}
Different surveys have been published regarding ICN and/or NDN, focusing on the general architecture~\cite{ahlgren2012survey, saxena2016named, ullah2018information}, specific components/features~\cite{bari2012survey, din2018caching, zhang2013caching, tyson2012survey, abdullahi2015survey, shang2017survey}, or well-defined applications and use cases~\cite{khelifi2019Named, nour2019survey}. However, there is a lack of surveys that are targeting access control in NDN. To justify the need and contribution of this survey, we summarize the existing surveys in Table~\ref{tab:related_surveys}. 

Ambrosin \textit{et al.}~\cite{ambrosin2018security} provide analysis on security and privacy features in Future Internet architectures designed by the U.S. National Science Foundation (NSF) including Nebula, MobilityFirst, Named Data Networking, and eXpressive Internet Architecture.
Tourani \textit{et al.}~\cite{tourani2018security} reviewed the security and privacy issues in ICN. The authors also covered access control schemes, but in a rather high-level manner, without focusing on the NDN architecture. Finally, they classified the existing solutions according to features such as naming, routing, caching, etc.
Similarly, Abdallah \textit{et al.}~\cite{abdallah2015survey} survey security attacks that are related to ICN implementations or may have impact to ICN features. In this study, they classified the existing attacks based on ICN aspects (i.e. naming-related attacks, routing-elated attacks, caching and other related attacks). The authors also discussed the security requirements to provide confidentiality, integrity, availability, and privacy. However, the authors did not cover access control solutions.
Lutz \textit{et al.}~\cite{lutz2016security} compare the CCN communication model with the current Internet model from the security and privacy point-of-view. The authors also present the security and privacy benefits of CCN, such as verifiable integrity, absence of device addressing, and protection against DoS attacks. Although they discuss some of the existing challenges and provide research directions, the authors do not offer a comprehensive survey on the existing security solutions and do not cover the access control aspect.
Similarly, Ghali \textit{et al.}~\cite{ghali2017encryption} assess the privacy of CCN. The authors discuss the existing privacy attacks and evaluate them using a custom CCN simulator.

On the other hand, Bouk \textit{et al.}~\cite{bouk2018named} discuss the security attacks and vulnerabilities in vehicular cyber-physical systems, and highlight a bunch of issues and challenges. Based on this study, the authors introduce an NDN-based cyber-resilient architecture to detect attacks and provide a resilience system. However, the authors did not cover access control aspects either in their study or in the proposed architecture.
Similarly, Khelifi \textit{et al.}~\cite{khelifi2018security} review the existing vehicular network attacks and classify them based on an NDN point-of-view. The authors identify various issues and challenges that require future investigation by the research community. However, the work targets vehicular environments without providing details about access control.

Besides the aforementioned survey and study efforts, \textit{Information-Centric Networking Research Group} (ICNRG) -- an IRTF research group provides a set of RFCs and drafts on ICN and its applicability. In particular,
RFC 7927~\cite{rfc7927} describes the challenges and issues of ICN before being widely deployed on core networks.
RFC 7945~\cite{rfc7945} provides an overview of the existing tools to evaluate ICN network focusing on the security aspects (e.g., authentication, authorization, access control, and logging), while ICN deployment guidelines have been provided to the broader community~\cite{irtf-icnrg-deployment-guidelines-05}.

\subsection{Motivation and Main Contributions}
Contrary to existing surveys, our work focuses on access control solutions in NDN. In comparison to~\cite{tourani2018security} (where authors study security, privacy, and access control from a bird's eye view), we thoroughly investigate and review the existing access control schemes in ICN/NDN\footnote{Without loss of generality, we use the terms `ICN', `NDN', and `CCN' interchangeably in this survey.}. We further classify existing solutions into two broad categories, and then each category into a set of sub-categories.

In this regard, the major contributions of our work are the following:
\begin{itemize}
	\item  We review content-based security, cryptographic algorithms and security protocols, and access control in NDN.
	\item We provide a comprehensive survey of the access control schemes (both encryption-based and encryption independent solutions) in NDN.
	\item We identify future challenges and research directions for access control in NDN. 
\end{itemize}

\subsection{Methodology of Survey on Access Control Mechanisms in Named Data Networking}

In the following, we present the methodology that we utilized to survey access control mechanisms in NDN~\cite{fei2020security}. We also present the taxonomy of existing literature.

\subsubsection*{Adopted Methodology}
\begin{figure}[!t]
	\centering
	\subfloat[Papers published per year.]{
		{\includegraphics[width=0.3\linewidth]{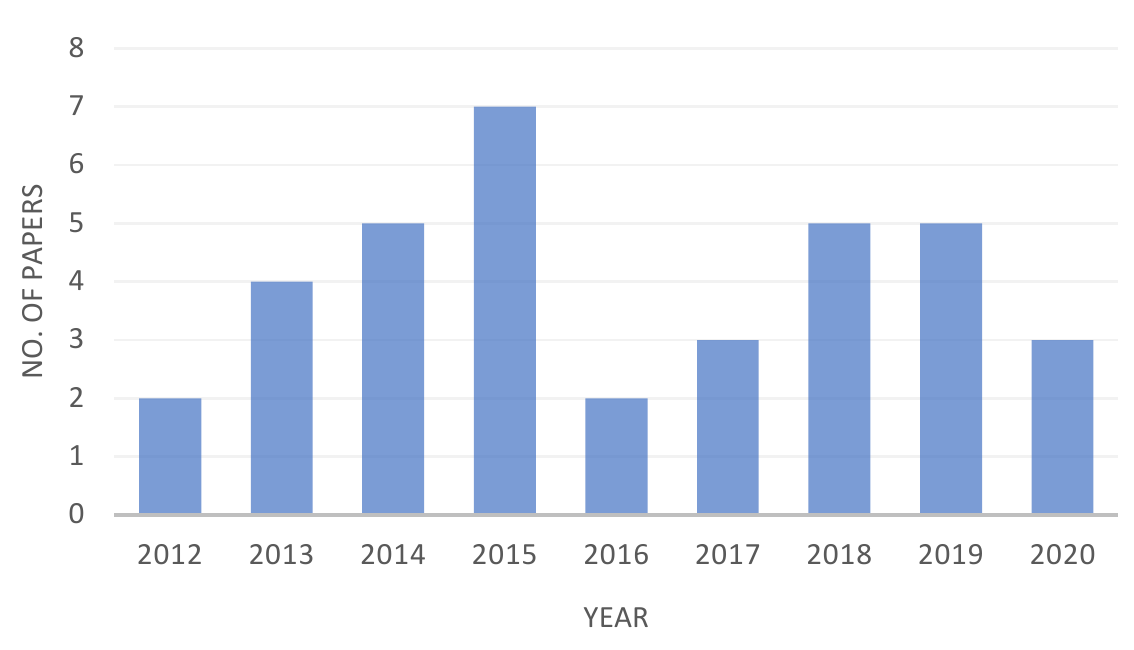}}
		\label{fig:statistic_1}
	}
	\hspace{3mm}
	\subfloat[Distribution of collected papers per database.]{
		{\includegraphics[width=0.3\linewidth]{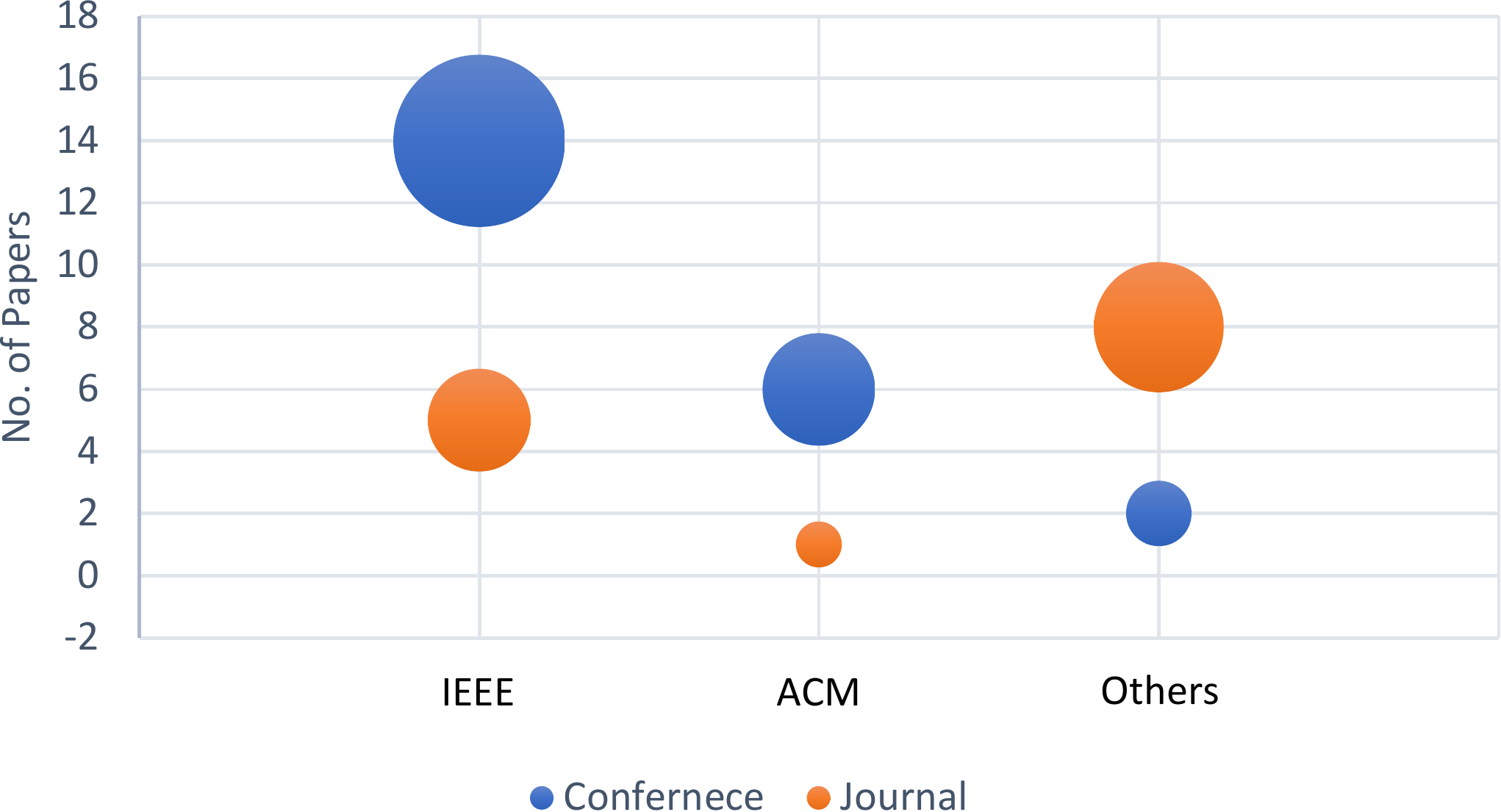}}
		\label{fig:statistic_2}
	}
	\hspace{3mm}
	\subfloat[Distribution of research papers per access control category.]{
		{\includegraphics[width=0.3\linewidth]{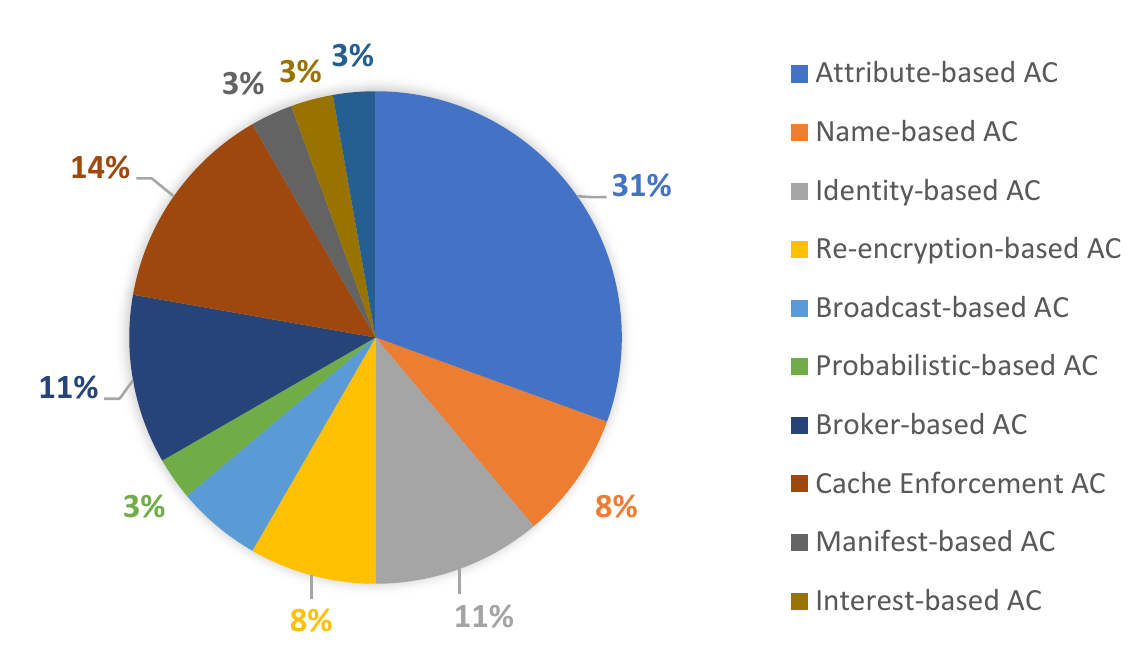}}
		\label{fig:statistic_3}
	}
	\caption{Statistic on collected papers.}
	\label{fig:statistics}
\end{figure}

To present a comprehensive survey of the state-of-the-art on access control mechanisms in NDN, we have adopted a systematic survey methodology. This methodology aims to provide a holistic overview described in the following steps:
 
\begin{itemize}
    \item \textbf{Surveyed Databases:} To achieve our objective, we have collected various published papers up to June 2020 available on domain-relevant electronic databases, including \textit{ACM Digital Library}, \textit{IEEE Xplore}, \textit{Science Direct}, and \textit{Springer Link}. Furthermore, we have collected articles related to this domain from \textit{arXiv}, \textit{Hindawi}, \textit{MDPI}, and \textit{Google Scholar} databases, as well as standardization research work from various \textit{IETF} working groups.
    
    \item \textbf{Targeted topic:} In our work, the targeted topic was ``\textit{Access Control in Named Data Networking}".
    
    \item \textbf{Search strings:} We adopted an automated search process and used the following keywords for our search: (``\textit{All Metadata}": ``\textit{Access Control Mechanism}" AND (``\textit{Named Data Networking}" OR ``\textit{Content-Centric Networking}")).
    
    \item \textbf{Filtering:} The filtering phase was separated into two steps: (i) we analyzed information related to each paper's metadata, such as title, abstract, and keywords; and (ii) we performed a full analysis of the paper.
    
    \item \textbf{Selection criteria:} This phase consisted of evaluating the papers retained from the previous phase by keeping only relevant papers and excluding those which are irrelevant, duplicated, and/or not written in English. The considered \textit{Inclusion Criteria} (IC) were: research papers focusing on the access control in NDN; while the considered \textit{Exclusion Criteria} (EC) were: papers that do not address access control in NDN as their main contribution as well as those consisting solely of bibliography, table of contents, references and keynote talks, editorial articles, or summaries of conferences. 
    
    \item \textbf{Research Questions (RQ):} For the retained papers, the following questions were considered: RQ1 -- How are the publications related to access control in NDN distributed over the years?, RQ2 -- Which are the major journals and conferences that publish articles about access control in NDN?
    
    In response to RQ1, Fig.~\ref{fig:statistic_1} shows that the research community started exploring access control solutions in NDN around 2012 (CCN was proposed in 2009 and NDN in 2010). Since then, the curve has taken an ascending trend. In 2016, fewer papers were published on the topic, but then researchers started concentrating on access control again as NDN was adopted as the communication enabler for different use-cases. These statistics confirm the relevance of the topic, which increasingly gains ground and interest within the research community. In response to RQ2, the bubble plot of Fig.~\ref{fig:statistic_2} summarizes the distribution of the collected papers per publication type (journal or conference) and database (ACM, IEEE, or other publishers).
\end{itemize}

\subsubsection*{Taxonomy of the Reviewed Literature}
A high-level analysis of the collected articles in the literature focusing on access control in NDN can be classified into two main categories: \textit{Encryption-based Access Control} and \textit{Encryption-independent Access Control}. Each category has a set of sub-categories. Figure~\ref{fig:statistic_3} shows the distribution of research papers.

\begin{figure}[!t]
	\centering
	\includegraphics[width=.9\linewidth]{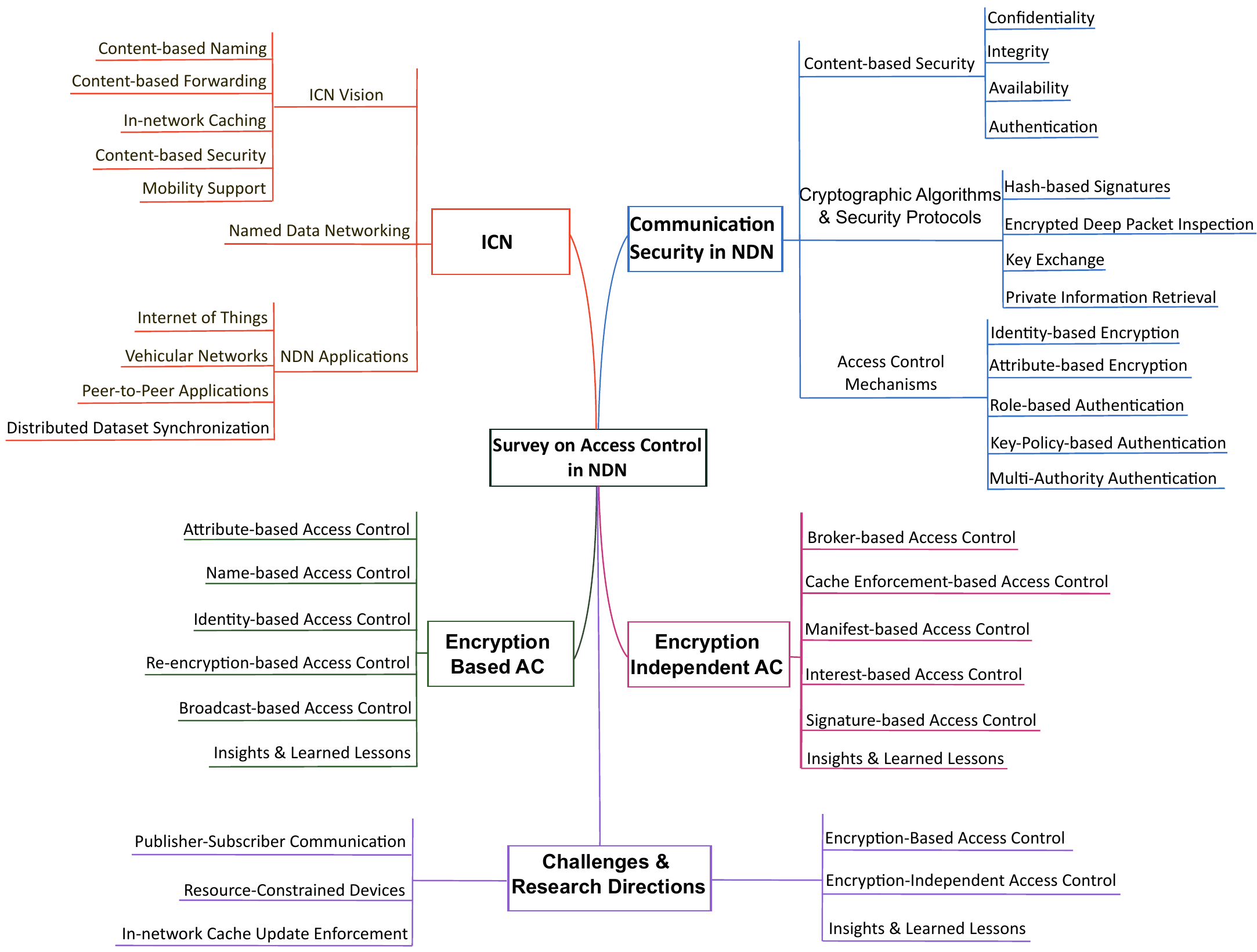}
	\caption{Structure of the paper and taxonomy.}
	\label{fig:classification}
\end{figure}

\subsection{Survey Structure}
The rest of this survey is organized as illustrated in Figure~\ref{fig:classification} and described as follows.  Section~\ref{sec:icn} introduces ICN and its features, as well as presents an overview of NDN.
Section~\ref{sec:cbs} discusses the communication security in ICN and the shift from channel-based security towards content-based security, presents the existing cryptographic algorithms and security protocols in NDN, and introduces the access-control concept as well as highlights the required security service.
Section~\ref{sec:enc-based} reviews the existing encryption-based access control.
Similarly, Section~\ref{sec:enc-ind} presents a review on encryption independent access control.
Section~\ref{sec:directions} presents challenges and highlights different guidelines and future research directions.
Finally, Section~\ref{sec:conclusion} concludes the survey.

\section{Information-Centric Network: A Bird's Eye View}
\label{sec:icn}
Information-Centric Networking~\cite{ahlgren2012survey} has been proposed as an alternative network architecture to address the major challenges in the existing IP-based network including routing, content sharing, scalability, and security. In the following, we provide an overview on ICN highlighting its global vision. Then, we present NDN architecture and its working principles as well as NDN applications.

\subsection{ICN Vision}
\begin{figure}[!t]
	\centering
	\includegraphics[width=.7\linewidth]{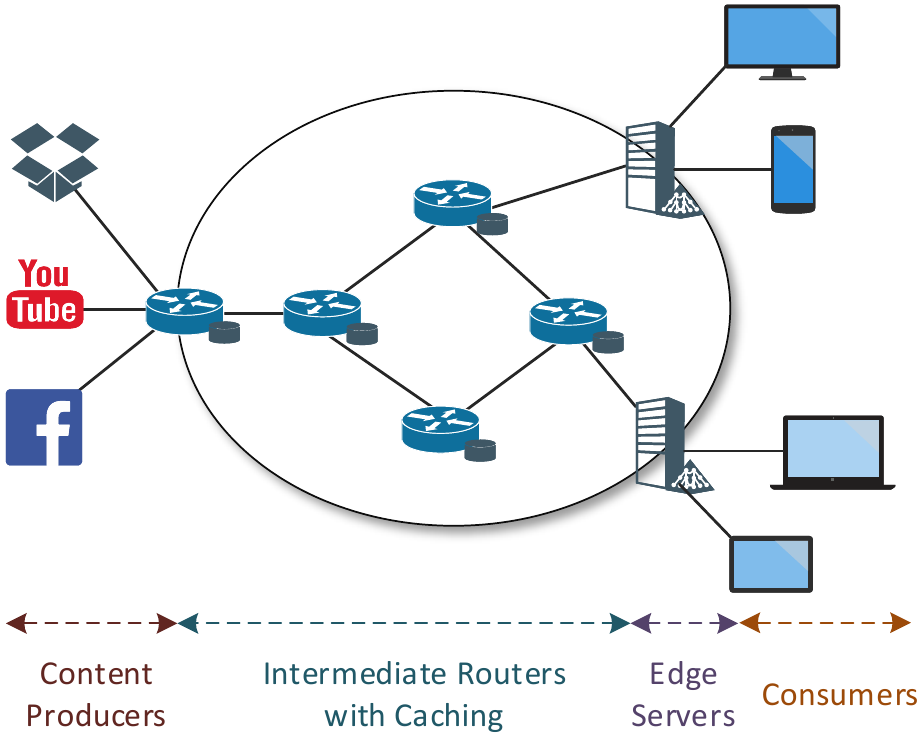}
	\caption{Basic ICN communication model.}
	\label{fig:icn_model}
\end{figure}

The ICN paradigm transforms the way the current Internet functions by leveraging content naming as the communication cornerstone, thus, replacing the current host-centric communication model~\cite{vasilakos2015information}. Figure~\ref{fig:icn_model} illustrates the basic communication model in ICN that includes a content producer (original publisher), intermediate routers with caching capabilities, consumers, and edge servers. 
We present an overview of the main ICN features below.

\subsubsection{Content-based Naming}
The content name~\cite{nour2017m2hav, nour2020hicniot} is the essential element in ICN. A name is used by the network as the content identifier, while it should also be persistent to validate the content. The used naming scheme must be scalable and allow for name aggregation and fast lookup performance. Four main types of naming schemes have been proposed in ICN:

\begin{itemize}
	\item \textit{Hierarchical names} are made of a set of components to identify the application, describe the service or the content. The structure of a hierarchical name is similar to \textit{Uniform Resource Identifiers} (URIs). Hierarchical names are usually user-friendly and convey semantic meaning to users. The hierarchical naming scheme enhances the network scalability since the name prefix can be aggregated and hence accelerate the lookup process.
	
	\item \textit{Flat names} are generally produced by applying hash algorithms to content. Flat names have neither a semantic meaning nor a structure. Therefore, the name is not human-friendly and hardly assigned to dynamic contents that are generated by demands. Flat naming cannot support routing aggregation, hence their scalability is an open issue.
	
	\item \textit{Attribute-Value based names} have a collection of attributes, in which each attribute has a name, a type, and a set of possible values (e.g., creation date, content type, version.). These attributes describe a single content piece along with its properties. Attribute-value based naming provides a mechanism for easy search operations through keywords. However, ensuring name uniqueness is challenging as one search request may lead to multiple results.
	
	\item \textit{Hybrid names} combine at least two of the previous schemes. The overall idea is considering features provided by different schemes to improve network scalability and performance, and heighten security and privacy. For example, taking advantage of name aggregation to enhance lookup operations, the fixed length of flat names to improve memory consumption, and attribute values for keyword-based search operations.
\end{itemize}

\subsubsection{Content-based Forwarding}
By adopting name instead of host address to identify content, ICN uses name-based routing~\cite{bari2012survey} to discover and deliver content to the requester. Due to the receiver-driven design, a consumer triggers a request for a specific content just by specifying its name. The request is forwarded hop-by-hop using a forwarding/routing table and the requested content name to find a cached version of the content in the network or reach the original producer. When the content is found, the data delivery process takes place to send the content back to the requester.

\subsubsection{In-network Caching}
Due to the use of location-independent and self-contained data packets, ICN enables in-network content caching~\cite{din2019puc} during communication.  The network becomes aware of content discovery and delivery~\cite{mastorakis2018real, shannigrahi2018scari}. In the ICN-based network, each node has the potential to cache content at the local cache-store and satisfy future demands. Therefore, the network witnesses massive improvements in terms of communication delay, content retrieval process, and quality of service.

\subsubsection{Content-based Security}
Due to the use of naming abstraction and in-network content caching, ICN adopts content-based security~\cite{tourani2018security}, in which security-related mechanisms are applied to the content itself rather than the communication channel/session. Different trust models have been introduced based on network services~\cite{kapetanidoureputation, khelifi2020blockchain, pi2018secure, fotiou2016decentralized, tschudin2016trust}. Besides, each data packet in ICN can be authenticated, while security-related information is bound to the content (e.g., the publisher's public/secret keys and signature).

\subsubsection{Mobility Support}
From the ICN perspective~\cite{tyson2012survey}, the content is decoupled from time and space. The content name is the only element used to discover and deliver it back to the consumer. When an ICN node moves from a network to another, it can re-issue any unsatisfied requests, and the producer or replica-node replies with the requested content without the need to request a new address during the movement. However, in this case we face two major issues: How the requested data can be delivered to a mobile consumer since no addresses are used, and how a consumer can request a data from a mobile producer? Another challenge arises in the case of producer mobility, specifically, how content can be retrieved from mobile producers~\cite{zhang2018kite, nour2020internet, serhane2020label}.

\begin{figure}[!t]
	\centering
	\subfloat[Data Discovery.]{
		{\includegraphics[width=0.47\linewidth]{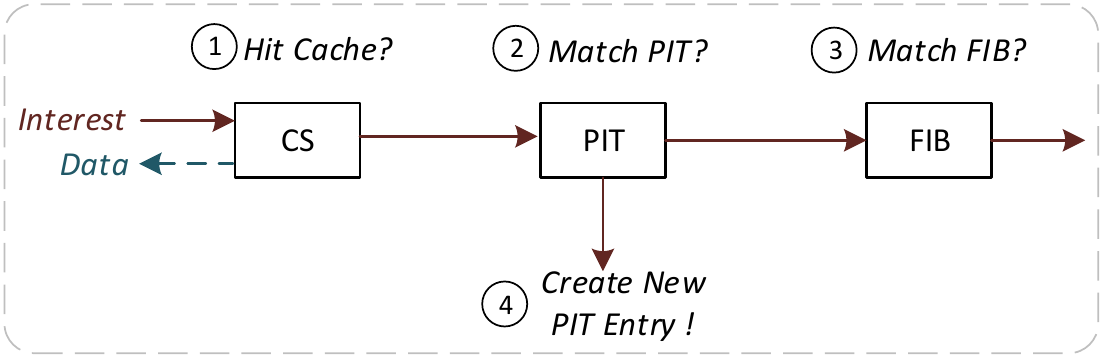}}
		\label{fig:interest_forwarding}
	}
	\hspace{3mm}
	\subfloat[Data Delivery.]{
		{\includegraphics[width=0.47\linewidth]{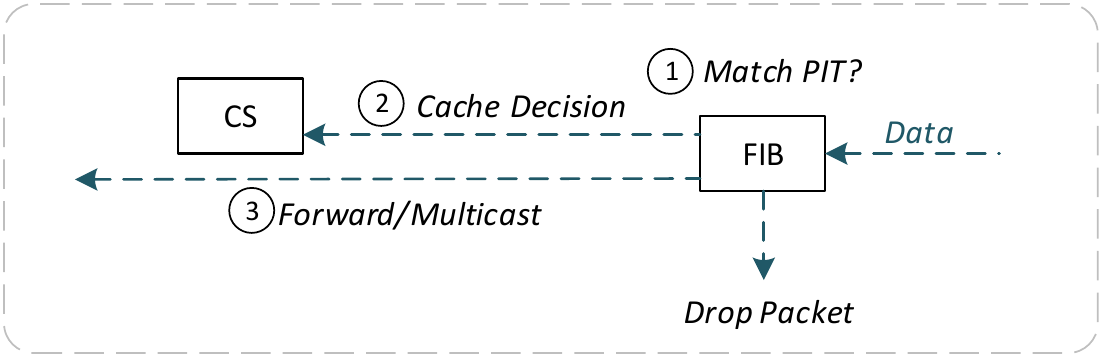}}
		\label{fig:data_forwarding}
	}
	\caption{NDN forwarding process.}
	\label{fig:ndn_forwarding}
\end{figure}

\subsection{Named Data Networking (NDN)}
NDN~\cite{zhang2010named} is the most prominent realization of the ICN vision~\cite{saxena2016named}. NDN uses two types of packets: Interest and Data. Interest packets are triggered from a consumer, in the form of a request message to ask for content from the network. The provider node or any replica node can reply with a Data packet that carry the requested content toward the consumer~\cite{li2018packet}. Both Interest and Data packets carry the name of the requested content. NDN is a named-based network where the routing plane is done by using names instead of IP addresses. NDN implements hierarchical, human-readable, and structured names. Moreover, NDN maintains three data structures: Content Store (CS), Pending Interest Table (PIT) and Forwarding Information Base (FIB).
Figure~\ref{fig:ndn_forwarding} illustrates the forwarding engine in Named-Data Network that can be divided in Data Discovery and Data Delivery:

\textbf{Data Discovery:}
After receiving an interest packet, the forwarding node consults its local CS to check if a copy of the requested content exists. If a match is found in the CS, then a data packet is sent out via the same interface that the interest packet was received. Otherwise, a PIT lookup is performed to verify if the same interest has been already forwarded. If a PIT match is found, the node adds the received interface ID to the PIT entry (also referred to as interest aggregation) and discards the interest packet. Subsequently, the FIB table is consulted to find the most suitable interface to reach the requested content. If a match is found, a new PIT entry is created using the requested content name and the ID of the received interface, and the interest packet is forwarded upstream via one or more outgoing interface(s) according to FIB~\cite{chan2017fuzzy, mastorakis2018experimentation}. 


\textbf{Data Delivery:}
In contrast to the interest forwarding process, only the PIT is involved in data forwarding. When a node receives a data packet, it checks the PIT to verify whether the associated request has been already forwarded via the node that receives the interest packet. If no match is found in the PIT table, it means that no interest carrying this name has been forwarded and thus the data packet is considered unsolicited and is dropped. Otherwise, the node forwards the data  to all listed interfaces in the PIT entry (multicast), and expunges the PIT entry. In the meantime (according to the current caching placement strategy), the data may or may not cached in the local CS.

\begin{figure}[!t]
	\centering
	\subfloat[Interest Packet.]{
		{\includegraphics[width=.21\linewidth]{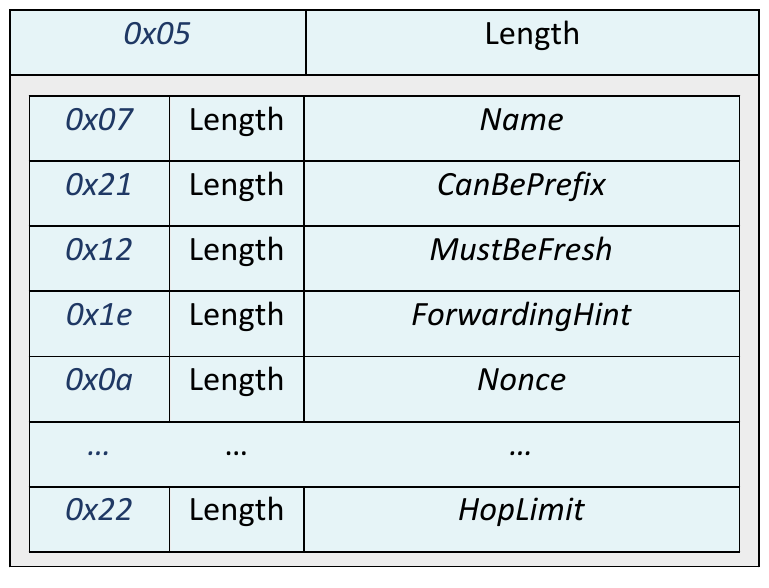}}
		\label{fig:interest_packet}
	}
	\subfloat[Data Packet.]{
		{\includegraphics[width=.75\linewidth]{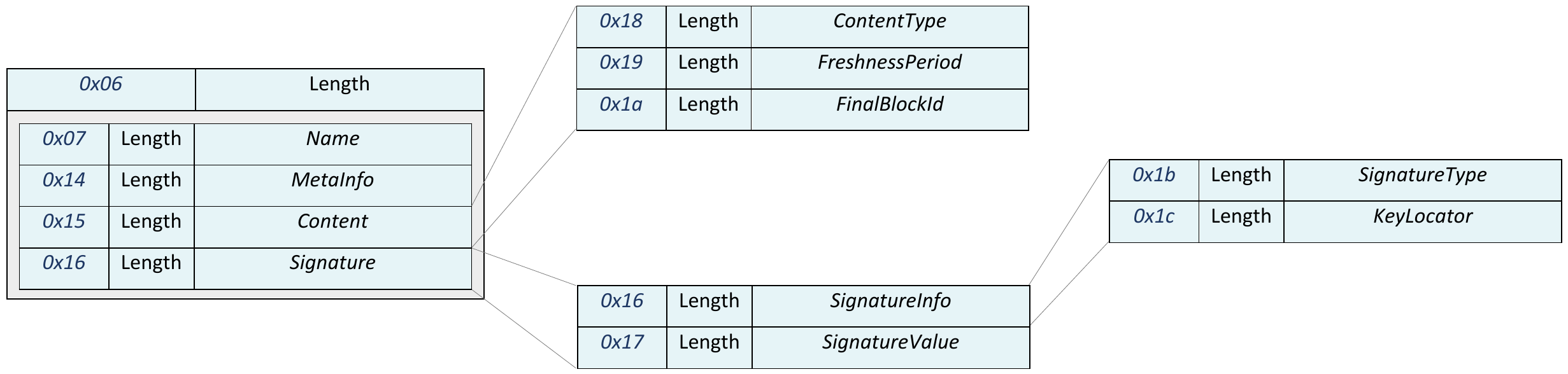}}
		\label{fig:data_packet}
	}
	\caption{Interest and Data packet structure.}
	\label{fig:packet_structure}
\end{figure}

\subsection{NDN Applications}
NDN is a receiver-driven architecture that can be integrated as a communication enabler for almost all kinds of today's applications and use cases~\cite{liang2018ndnizing}.
Below, we discuss existing efforts on areas that can benefit from access control mechanisms, such as the Internet of Things, Vehicular Networks, peer-to-peer applications, and distributed dataset synchronization protocols.

\subsubsection{Internet of Things}
In the coming years, there will be billions of connected smart devices such as sensors, smartphones, cars and data centers, deployed in any environment with computation capability and interoperable interconnection, exchanging information online. These form of connection is known as the Internet of Things (IoT)~\cite{al2015internet} that will be deployed with sensing and intelligence capabilities so that they can not only communicate, but collecting data, negotiate, collaborate and exchange the collected values.

Deploying IoT applications of top of ICN~\cite{nour2019survey, mastorakis2020icedge, mastorakis2019towards, nour2020compute} is promising to enhance the overall network performances. Indeed, most of information produced by IoT smart devices can be regarded as content~\cite{nour2019hicniot}, consumers in the networks request data in IoT context without the need of the location of the sensors or the actuator, no end-to-end session is required for content retrieval~\cite{nour2017aal}, where ICN target the content in the network by its name rather than its address. Also, most of IoT devices request the same content in the network such as asking for humidity value for a specific place, or query some information, or data monitoring~\cite{nour2018nncp}. In the second hand, we find out that ICN nodes can act as node-replica by using content store, and serve for future requests regardless of the content reachability from the original source, that improve the data retrieval, and reduce the latency~\cite{nour2020distributed, nour2018ncp, khelifi2020effectiveness, serhane2020cns}. In such scenarios, NDN is more suitable for IoT than IP, not only for the rapid content delivery, but also for its receiver-driven design, and request aggregation~\cite{khelifi2018bringing, mastorakis2019towards, krol2019compute}.

\subsubsection{Vehicular Networks}
Recently, substantial efforts and funding projects have been shown in the area of Vehicular Ad hoc Networks (VANETs) \cite{rettore2018vehicular, mejri2014survey}. The main purposes of vehicular networks and applications are to provide a comfortable life, efficient transportation \& safety services, and secure data sharing. However, the characteristic of vehicular communication is more challenging especially with the high and dynamic movement of vehicles that affects the network topology and the reliability of communication \cite{mejri2016gdvan, mokdad2015djavan, vusirikala2016hop}.

Towards the merger of NDN with VANET networks~\cite{khelifi2019Named}, various research efforts have been proposed to enhance the core NDN architecture or target different features and aspects~\cite{ahmed2017named, hussain2018realization, khelifi2018reputation}. The use of content name to identify services and content may help to provide more independent communication, especially when mobility factor is taken into consideration~\cite{khelifi2019NHE, khelifi2019LQCC, nour2020collaborative}. Coupling the naming with in-network caching is also promising to improve the overall network performance and providing an efficient content delivery~\cite{khelifi2018optimized, ostrovskaya2018towards}. A mobile node is not required to get a new IP address when move from a network to another. Indeed, it is only required to send a request packet specifying the required name. Doing so, the mobility issue can be enhanced.

\subsubsection{Peer-to-Peer Applications}
Peer-to-peer applications, such as BitTorrent for file sharing, have faced several challenges running on top of TCP/IP. Among others, such challenges include peer discovery, peer selection, and traffic localization~\cite{mastorakis2017ntorrent, mastorakis2020dapes, mastorakis2019peer}. In NDN, the content is decoupled from the location of its production and as a result, explicit peer discovery and selections mechanisms are not needed since the NDN forwarding plane will retrieve the requested data from the closest peer that can offer it. Moreover, in-network caching and Interest aggregation can effectively mitigate flash-crowd scenarios, a common problem with TCP/IP-based peer-to-peer file sharing applications. Such applications capitalize on data sharing, therefore, access control mechanisms may be useful to ensure that only authorized peers have access to certain pieces of data.

\subsubsection{Distributed Dataset Synchronization Applications}
NDN transforms the problem of synchronizing datasets in a distributed manner to a problem of namespace synchronization. Several synchronization protocols have been proposed both for infrastructure-based~\cite{fu2015synchronizing, de2017design, fu2014isync} and infrastructure-less environments~\cite{lidistributed}. In this use-case, access control mechanisms can be used to determine which entities can join a synchronization group and get access to the synchronized dataset.

\section{Communication Security in NDN}
\label{sec:cbs}
Indeed, the use of content names instead of host addresses allows the network layer to decouple the content from its original location, consequently enabling in-network caching. This feature promises to distribute the content towards different locations~\cite{marxer2017schematized}, usually close to consumers, in order to decrease the delay, improve the quality of service~\cite{khelifi2019qos}, and cover a large number of consumers regardless of the availability of the original content producer. In such cases, the original producer loses the control over its content including security, privacy, and access control~\cite{nour2019sp, boussada2019pp}.

\subsection{Content-based Security}
Traditional security protocols~\cite{volkova2019security} such as Transport Layer Security (TLS) and Secure Sockets Layer (SSL) may not be a good fit for NDN as they are based on building a secure channel or session between the content producer and the consumer. However, in NDN, content may be delivered from different locations and content stores; some content can be delivered from the network edge nodes, while other may be satisfied by neighboring nodes or intermediate nodes at the core network (i.e., ISP -- Internet Service Provider)~\cite{tsudik2016ac3n, chaabane2013privacy, bernardini2019privicn}. 

Content-based security concept tends to provide the main security and privacy services~\cite{benmoussa2020msidn} such as:

\begin{itemize}
	\item \textit{Confidentiality}: Confidentiality refers to the task of protecting the content from being accessed by unauthorized users. Only the users who are authorized to do so can gain access to sensitive data. If the system fails to maintain confidentiality, a user who should not have access has the ability to get it.
	
	\item \textit{Integrity}: Integrity refers to the task of ensuring the authenticity of content (i.e., the content itself or the producer of the content have not been forged). If the system fails to provide integrity, anyone in the network can provide content and pretend that they re the owners of the information.
	
	\item \textit{Authentication}: Authentication refers to the process of ensuring and confirming a user's identity. If the system fails to provide authentication, anyone in the network can access any content regardless of their identity or credentials.
	
	\item Availability: Availability refers to the task of ensuring that authorized parties are able to access the content when needed.
\end{itemize}

The content-based security concept aims to secure the content itself on a per packet basis rather than securing the communication channel~\cite{lutz2016security}. Fundamentally, NDN packets are encoded using a Type-Length-Value (TLV) format~\cite{tlv}. Figure~\ref{fig:packet_structure} illustrates both Interest and Data packet encoding. Each packet is a collection of TLVs, where a TLV may contain nested TLVs. NDN forwarders identify these packets using their type, while each TLV has a unique type. 

In NDN, each Data packet is secured and cryptographically signed~\cite{yang2019securing, rezende2018distributed}. The TLV \texttt{Signature} specifies all information related to the packet signature including: (a) \texttt{SignatureInfo}: to describe the signing algorithm, and relevant information about certificates and keys; and (b) \texttt{SignatureValue}: to represent the signature itself in bits. It is important to highlight that regular Interest packets are typically not signed; however, NDN introduces \textit{Interest Signature} to produce signed Interest packets. NDN supports a wide range of signatures such as \textit{DigestSha256}: using SHA-256 digest for data integrity that produces only 32 bytes; \textit{SignatureSha256WithRsa}: using an RSA signature over a SHA-256 digest which is an efficient asymmetric encryption method where we only need to verify that the data has not been tampered; \textit{SignatureSha256WithEcdsa}: using an ECDSA signature over a SHA-256 digest, however ECDSA signatures are not deterministic and they depend on a random number generator to generate the signatures that may have a different value after each signing operation, hence the correctness of the value of these signatures can only be tested by verifying with the public key; and \textit{SignatureHmacWithSha256}: using a SHA256 hash-based message authentication codes. The efficiency and performance of these signature algorithms vary and are mainly based on the used encryption algorithm.

Content retrieval in NDN shifts from the session-based security towards content-based security. It aims to provide different security and privacy services. In the next section, we discuss what kind of cryptographic algorithms, security protocols, and techniques can be used to provide the core network security.

\begin{table}[!t]
	\makegapedcells
	\centering
	\caption{Cryptographic algorithms \& security protocols in NDN.}
	\label{tab:crypto_algorithms}
	\scriptsize
	\begin{tabular}{p{0.08\textwidth}p{0.24\textwidth}p{0.11\textwidth}cccp{0.17\textwidth}}
		\toprule
		\makecell[c]{\textbf{Approach}} &
		\makecell[c]{\textbf{Features}} &
		\makecell[c]{\textbf{Target}} &
		\makecell[c]{\textbf{Level}} &
		\makecell[c]{\textbf{Storage}} &
		\makecell[c]{\textbf{Computation}}&
		\makecell[c]{\textbf{Drawbacks}} \\
		\midrule
		
		Hash-based Signatures&
		$\bullet$ Ensure integrity and authentication \newline
		$\bullet$ RSA, ECDSA, and MSS \newline
		$\bullet$ High and fast signature generation \newline
		$\bullet$ Small-sized private and public keys \newline &
		$\bullet$ Integrity \newline
		$\bullet$ Authentication &
		C, P &
		H&
		M&
		$\bullet$ One-time signatures \newline
		$\bullet$ Signature size \newline \\
		
		DPI &
		$\bullet$ Ensure content privacy \newline
		$\bullet$ Inspect packets \newline 
		$\bullet$ DPI, Blind-Box \newline &
		$\bullet$ Integrity &
		R &
		M &
		H &
		$\bullet$ Violate name and user privacy \newline
		$\bullet$ Select trusted router(s) \newline \\
		
		Key Exchange&
		$\bullet$ Allow the use of cryptographic algorithms \newline
		$\bullet$ Session-based communication \newline
		$\bullet$ Diffie-Hellman \newline &
		$\bullet$ Integrity \newline
		$\bullet$ Authentication &
		C, P &
		H &
		M &
		$\bullet$ One-to-many communication \newline
		$\bullet$ Many-to-many communication \newline \\
		
		PIR &
		$\bullet$ Prevent the host from identifying requested content. &
		$\bullet$ Integrity &
		R &
		L &
		H &
		$\bullet$ Design and implementation complexity 
		\\
		
		\bottomrule
	\end{tabular}
	\caption*{\it \scriptsize * C: Consumer, P: Producer, R:Router | L: Low, M: Medium, H: High}
\end{table}

\subsection{Cryptographic Algorithms \& Security Protocols in NDN}
\label{sec:algo_proto}
Most existing cryptographic algorithms and security protocols can be applied to NDN with slight or more extended modifications. This section provides a comprehensive overview of such algorithms and protocols, Table~\ref{tab:crypto_algorithms} summarizes our discussion.

\subsubsection{Hash-based Signatures}
Hash-based signatures~\cite{butin2017hash} ensure the integrity and authentication of NDN data packets. Various hash-based signing algorithms can be used in NDN such as RSA and ECDSA, which are characterized by high and fast signature generation and verification with private and public keys. Moreover, these signatures are usually based on an One-Time Signature (OTS) system~\cite{shafieinejad2017post} that allows the use of key pairs to only sign one message securely. An N-time signature system is another way that allows signing $N$ messages securely.
The Merkle Signature Scheme (MSS)~\cite{buchmann2006cmss} is another hash-based, N-time signature system that uses an OTS system as an element. MSS is a binary tree that has a height $h$ that is used to create $2^h$ OTS key pairs. Each node in the tree has a value that represents the hash values of its child nodes. The public key is the value of the root node, while the private key is the combination of all OTS private keys with the index of the next OTS private key.

\subsubsection{Encrypted Deep Packet Inspection}
Deep Packet Inspection (DPI)~\cite{xu2016survey} can be used in NDN to inspect packets and check whether they contain sensitive data. This process violates content and user privacy. However, by encrypting the packet payloads, DPI cannot be performed, and the privacy services can be preserved. BlindBox~\cite{sherry2015blindbox} has been proposed in order to perform the DPI directly on encrypted payload. It is based on two main classes of DPI computation with different privacy assumptions and guarantees: exact match privacy and probable cause privacy. The first class includes DPI applications that depend on exact string matching, while the second class can support all DPI applications.
%
NDN uses plaintext names instead of IP addresses, which raises concerns over content privacy. Using encrypted or pseudo names may alleviate such concerns, but still, name obfuscation may interfere with name-based routing. Finally, how to select a trusted anonymizer for name obfuscation is an open question.

\subsubsection{Key Exchange}
Key exchange, also known as key establishment, is an important security aspect of asymmetric crypto-systems, so that keys between communicating parties are exchanged in order to be used in cryptographic algorithms. Diffie-Hellman~\cite{malik2019survey} is probably the most widely-used key exchange protocol. This algorithm assumes session-based (point-to-point) communication, thus requiring modifications for communication among multiple parties (one-to-many or many-to-many communication)~\cite{de2017design, li2018data, lidistributed}. Hence, a secure group-based protocol is more suitable for such scenarios~\cite{nour2019gbps}. The content producer may generate a new key pair (public/private keys) based on the number of active subscribers in the communication to secure the published content. It is important to highlight that the key exchange/generation protocol should be aware of the active subscribers, previous active subscribers should not be allowed to decrypt newly generated content. Also, whenever a new subscriber joins/leaves the subscription, the protocol should exchange/generate new key pairs~\cite{bian2013deploying}.

\subsubsection{Private Information Retrieval}
Private Information Retrieval (PIR) is a cryptographic primitive widely used in databases. It aims to prevent the database server or the host service from identifying which record has been requested. This concept is important in ICN/NDN. Related work has investigated private lookup operations for named data~\cite{tschudin2016private}, essentially hiding information about the names of the data packets that were retrieved.


\subsection{Access Control Mechanisms}
Authentication and access control are important security aspects, especially with the increase in the devices that generate content. To this end, various access control solutions have been proposed in the literature. We further discuss these concepts including their pros and cons below. Table~\ref{tab:ac_mechanisms} summarizes this discussion.

\begin{table}[!t]
	\makegapedcells
	\centering
	\caption{Summary of access control mechanisms.}
	\label{tab:ac_mechanisms}
	\scriptsize
	\begin{tabular}{p{0.12\textwidth}p{0.24\textwidth}ccccp{0.2\textwidth}}
		\toprule
		\makecell[c]{\textbf{Approach}} &
		\makecell[c]{\textbf{Features}} &
		\makecell[c]{\textbf{Level}} &
		\makecell[c]{\textbf{NDN Applicable}} &
		\makecell[c]{\textbf{Storage}} &
		\makecell[c]{\textbf{Computation}} & 
		\makecell[c]{\textbf{Drawbacks}} \\
		\midrule
		
		Identity-based Encryption&
		$\bullet$ Generates public keys from known unique identifiers \newline 
		$\bullet$ Does not require distribution public keys &
		C, P &
		\xmark& 
		M  &
		L &
		$\bullet$ Third party authentication \newline
		$\bullet$ Requires a centralized server\newline
		$\bullet$ Requires a secure channel\newline 
		\\
		
		Attribute-based Encryption &
		$\bullet$ Uses a set of attributes \newline
		$\bullet$ Provides fine-grained access control&
		C, P &
		\cmark &
		L&
		L&
		$\bullet$ Challenge of attribute revocation\newline 
		\\
		
		Role-based Authentication&
		$\bullet$ Use of the role concept \newline
		$\bullet$ One-to-many communication model \newline
		$\bullet$ Key revocation &
		P &
		\xmark &
		H &
		H &
		$\bullet$ Additional overhead\newline
		$\bullet$ Not suitable for dynamic network conditions \newline 
		$\bullet$ Third party authentication\newline 
		\\
		
		Key-Policy-based  Authentication &
		$\bullet$ Provides fine-grained access control\newline
		$\bullet$ Determines which attributes to use with content &
		C, P &
		\cmark &
		H &
		H &
		$\bullet$ Additional overhead \newline 
		$\bullet$ Third party authentication 
		\\
		
		Multi-Authority  Authentication &
		$\bullet$ Uses attributes from different authorities \newline
		$\bullet$ Uses symmetric encryption \newline
		$\bullet$ Expressive and efficient scheme &
		C, P &		
		\xmark &
		H &
		H &
		$\bullet$ Additional overhead 
		\\
		
		\bottomrule
	\end{tabular}
	\caption*{\it \scriptsize * C: Consumer, P: Producer | L: Low, M: Medium, H: High}
\end{table}

\subsubsection{Identity-based Encryption}
Identity-based encryption is an AC technique based on public key infrastructure. A private key generator is used to generate master public and private keys. A user is able to decrypt the content by getting the master private key after getting authenticated by the private key generator through the user identity. This scheme aims at reducing the complexity of the encryption process, however the whole system is based on trusting the third party that generates the master keys and authenticates users.

\subsubsection{Attribute-based Encryption}
In Attribute-based Encryption, an attribute authority generates public and master private keys for a content producer. The content owner uses a set of attributes to allow authorized users to access content as well as generates the keys subsequently used to encrypt content. A content consumer uses his/her own private key to decrypt the content if the attributes are matched. Although this scheme provides fine-grained access control, the attribute authority needs to use the public key of each authorized user to generate keys and encrypt content.

\subsubsection{Role-based Authentication}
In role-based authentication, the content owner uses the role concept to authenticate users and encrypt the content. A role is assigned based on the responsibilities and qualifications of an entity. An authenticated user has the privileges to access content according to the assigned role. A third-party entity or the content owner can manage the responsibility to assign roles, and consequently revoke a role if a user is no longer authorized.  This scheme is suitable for one-to-many communication model and easy to implement and handle key revocation; however, it may result in overhead in cases of dynamic changes of the users joining and leaving the communication group.

\subsubsection{Key-Policy-based Authentication}
In key-policy-based authentication, the encrypted content is attached with a set of attributes. The private key issued by a third party is associated with an access policy structure that describes the user's identity. A user can decrypt content only if the access policy in his private key satisfies the attributes attached with the encrypted content. Although this scheme may provide fine-grained access control, deciding which attributes to use, attaching them to the content and using an extra access control structure produce overhead and may not scale well.

\subsubsection{Multi-Authority Authentication}
Multiple authorities issue attributes to users and assign an access policy structure to the content using attributes from different authorities.  
The content owner divides the content into different chunks in which each chunk is encrypted using symmetric encryption techniques, defines a set of access policies using attributes from multiple attribute authorities, and then encrypts the content. A user can decrypt the content only if his/her attributes satisfy the access policy associated with the content. However, managing the various authorities and attributes may result in considerable overhead.

The aforementioned schemes are widely used in today's IP-based networks. Migrating them to content-oriented networks require modifications especially due to the use of ubiquitous in-network caching and content naming. In the following sections, we review the existing NDN-based AC solutions. We broadly classify them into two main categories: encryption-based and encryption-independent AC solutions.

\subsection{Insights \& Learned Lessons}
NDN provides numerous advantages over the IP-based Internet such as efficient data dissemination, in-network caching, and content-based security. In fact, content-level security is promising to enforce security at the content itself rather than the communication channel. Authentication and access control rules can be applied to the content and not the provider. Hence, all rules and access policies need to be associated with the content during transmission and caching. 

Some of the key learned lessons that should be noted are the following:
\begin{itemize}
	\item In ICN/NDN, naming becomes a `\textit{shared}' layer between applications and the network. To this end, ICN transforms the problem of designing access control mechanisms to the design of appropriate naming schemes that determine the permissions of each entity for access to produced content. Naming also facilitates the realization of attributed-based access control mechanisms.
	\item ICN/NDN offers built-in security mechanisms directly at the network layer. Such mechanisms, apart from securing the retrieved content itself, can be used to secure the retrieval of access control rules, certificates, and encryption/decryption keys.
\end{itemize}

\section{Encryption-based Access Control}
\label{sec:enc-based}
In this section, we review existing encryption-based access control solutions in NDN. We refer to an encryption-based scheme as the process where the content producer encrypts the content and establishes access control rules. The content consumers must explicitly get authenticated by the publisher. The existing solutions have been further classified into subcategories based on the types of encryption. Table~\ref{tab:encryption_based_ac} summarizes the reviewed solutions.

\begin{table}[ht]
	\centering
	\caption{Summary of Encryption-based access control mechanisms.}
	\label{tab:encryption_based_ac}
    \tiny
	\begin{tabular}{cp{0.29\textwidth}
			cp{0.1\textwidth}cccp{0.26\textwidth}}
		\toprule
		
		\textbf{Ref.} &
		\makecell[c]{\textbf{Approach}} &
		\textbf{Level} &
		\textbf{Content} &
		\textbf{Cache} &
		\textbf{O/H} &
		\textbf{Comp.} &
		\makecell[c]{\textbf{Major Drawbacks}} \\
		\midrule
		
		\multicolumn{8}{c}{\textit{\textbf{Attribute-based Access Control}}} \\ \midrule
		
		\cite{ion2013toward}&
		$\bullet$ Attribute based encryption mechanism &
		C &
		Dynamic &
		\cmark &
		H &
		L &
		$\bullet$ No evaluation presented \newline
		$\bullet$ Missing revocation schemes \\
		
		\cite{li2014toward}&
		$\bullet$ Privacy preserving for published and cached content & 
		C, P &
		One-Time Usage Content &
		\cmark &
		L &
		L &
		$\bullet$ Relies on a trusted third party \\
		
		\cite{da2015access}&
		$\bullet$ Attribute based encryption with revoked privileges & 
		N &
		Dynamic &
		\cmark &
		H &
		M &
		$\bullet$ Third party authentication \newline
		$\bullet$ Single point of failure \newline
		$\bullet$ Proxy needs to be always online \\
		
		\cite{li2018attribute}&
		$\bullet$ Attribute-based encryption naming scheme & 
		R &
		Dynamic &
		\cmark &
		H &
		H &
		$\bullet$ Feasible only with flat names \newline
		$\bullet$ Cannot be applied to hierarchical names \\
		
		\cite{grewe2017encircle}&
		$\bullet$ Access control management based on attribute encryption & 
		N &
		Static &
		\cmark &
		H &
		H &
		$\bullet$ Third party authentication \newline
		$\bullet$ Component placement and communicate models are missing \\
		
		\cite{borgh2017employing}&
		$\bullet$ Use of central authority for attribute-based Encryption & 
		N &
		Dynamic &
		\cmark &
		L &
		H &
		$\bullet$ Use smaller policies \newline
		$\bullet$ Authority problem\\ 
		
		\cite{wu2019chtds} &
		$\bullet$ Ciphertext policy using attribute encryption. & 
		P &
		Static &
		\cmark &
		H &
		L &
		$\bullet$ Requires publisher to be always online. \\ 

		\cite{wu2020multi} &
		$\bullet$ Use of attribute authorization entity. & 
		C, P &
		Static &
		\cmark &
		H &
		H &
		$\bullet$ Relies on proxy security \\

		\cite{ramani2019ndn} &
		$\bullet$ Use of attribute-based signatures. & 
		C, P &
		Static &
		\xmark &
		H &
		H &
		$\bullet$ Costly verification process \newline
		$\bullet$ Large signature size \\ \midrule
		
		\multicolumn{8}{c}{\textit{\textbf{Name-based Access Control}}} \\ \midrule
		
		\cite{hamdane2013access}&
		Identity-based cryptography mechanism & 
		P &
		Dynamic &
		\cmark &
		H &
		M &
		$\bullet$ Clients can access previously published content \newline
		$\bullet$ Does not scale well \\
		
		\cite{wang2014session}&
		$\bullet$ Session-based mechanism &
		P &
		Dynamic &
		\xmark &
		H &
		H &
		$\bullet$ Replicated content in the network \\ 
		
		\cite{hamdane2013data} &
		$\bullet$ Data-based access control using encryption and lock password & 
		P &
		Static &
		\cmark &
		M &
		H &
		$\bullet$ Hard to generate/update access rules \\ \midrule
		
		\multicolumn{8}{c}{\textit{\textbf{Identity-based Access Control}}} \\ \midrule
		
		\cite{wood2014flexible}&
		$\bullet$ Combining proxy re-encryption with identity-based cryptography & 
		C, P &
		Static &
		\xmark &
		M &
		M &
		$\bullet$ Requires publisher to be always online \\
		
		\cite{fotiou2018rendezvous}&
		$\bullet$ Use of rendezvous points to enforce AC in pub-sub & 
		P &
		Static &
		\xmark &
		H &
		H &
		$\bullet$ Heavy and intensive computation \newline
		$\bullet$ Requires publisher to be always online \\
		
		\cite{hamdane2015credential} &
		$\bullet$ Access control using certified credentials & 
		P &
		Static &
		\xmark &
		M &
		M &
		$\bullet$ Use of a centralized entity \newline
		$\bullet$ Not efficient in large-scale networks \\
		
		\cite{tseng2018fgac} &
		$\bullet$ Fine-grained access control. & 
		C, P &
		Static &
		\xmark &
		H &
		H &
		$\bullet$ Trust-ability issues \\ \midrule
		
		\multicolumn{8}{c}{\textit{\textbf{Re-encryption-based Access Control}}} \\ \midrule
		
		\cite{mangili2015cache}&
		Update of access rules at content stores &
		P, R &
		Static &
		\cmark &
		H &
		H &
		$\bullet$ No guarantees in terms of updating all cached versions \newline
		$\bullet$ Necessitates the availability of original producer \\
		
		\cite{zheng2015achieving}&
		$\bullet$ Dual-phase encryption mechanism \newline
		$\bullet$ One-time decryption key, proxy re-encryption, all-or-nothing transformation &
		P, R &
		Static &
		\cmark &
		H &
		H &
		$\bullet$ Scalability is an open question \newline 
		$\bullet$ Necessitates the availability of original producer \newline
		$\bullet$ Trustability of edge routers \\ \midrule
		
		\multicolumn{8}{c}{\textit{\textbf{Broadcast-based Access Control}}} \\ \midrule
		
		\cite{misra2013secure}&
		$\bullet$ Broadcast decryption keys in the network &
		P, C &
		Static &
		\xmark &
		H &
		H &
		$\bullet$ Large number of broadcast keys \\
		
		\cite{misra2017accconf}&
		$\bullet$ Broadcast secure content at CS &
		P, R &
		Static &
		\xmark &
		H &
		H &
		$\bullet$ Dynamic changes of in-network content stores \\ \midrule
		
		\multicolumn{8}{c}{\textit{\textbf{Probabilistic-based Access Control}}} \\ \midrule
		
		\cite{chen2014encryption}&
		$\bullet$ Probabilistic encryption to prevent unauthorized access &
		P &
		Static &
		\xmark &
		H &
		H &
		$\bullet$ Requires publisher to be always online \newline
		$\bullet$ Impact of false positive errors \\ 
		
		\bottomrule
	\end{tabular}
	\caption*{\it \scriptsize * N: Network C: Consumer, P: Producer, R: Router | L: Low, M: Medium, H: High}
\end{table}

\subsection{Attribute-based Access Control}
Attribute-Based Encryption (ABE)~\cite{goyal2006attribute} incorporates public-key encryption where the secret key heavily depends on a set of attributes. Here, we differentiate two main types of ABE: (a) Key-Policy Attribute-Based Encryption (KP-ABE) \cite{li2019key}: the access tree is represented by a set of secret keys that define the privileged scope, and (b) Ciphertext-Policy Attribute-Based Encryption (CP-ABE) \cite{jiang2018ciphertext}: the access tree is represented by attributes in a ciphertext. The main advantages of ABE is that it is flexible and may provide access control enforcement to a group of users \cite{belguith2018phoabe}. However, it must be deployed based on a single and centralized authority, a fact that may affect the scalability of the system.
%

Ion \textit{et al.}~\cite{ion2013toward} propose an attribute-based encryption mechanism for content-centric data privacy in ICN by attaching the access control policies to the content itself
The client decrypts the content only if its symmetric key satisfies the access control policies that are included in the cipher-text or the key itself. However, the authors did not evaluate their solution in comparison with other proposed solutions.

Li \textit{et al.}~\cite{li2014toward} introduce a privacy-preserving content access control scheme for content already published and cached by other nodes. The proposed scheme is based on Attribute-Based Encryption and requires a trusted third party with the responsibility to assign attributes to other entities. Any node in the network has a unique identifier and a set of attributes. The content producer assigns a combination of attributes to the content before its publication to specify who can access it without explicitly knowing the consumer. It also generates a random symmetric key to encrypt the content. The consumer uses its attributes to access the content, if the attributes satisfy the encryption policy, the consumer then gets the random symmetric key, and consequently decrypt the content. The major limitation of this system is that it is based on a trusted third party.

Da \textit{et al.}~\cite{da2015access} design an attribute-based encryption scheme to provide fine-grained access policies in NDN with the ability to revoke privileges. A proxy node is used to provide access to secured data and inspect rules for revocation. Consumers register in the network based on a set of descriptive attributes that allow the proxy to authenticate users~\cite{suksomboon2017ipres}. The authors divide the content into two main parts: the content/data itself, and the structure to define access rules. The former can be cached by intermediate nodes, while the latter is cached and used only by the proxy node. Although this scheme may eliminate the necessity to connect with the original producer to fetch access rules, it allows proxy nodes to decrypt the content without any trust enforcement policies.

Li \textit{et al.}~\cite{li2018attribute} address the access control policies for distributed cached content in ICN. The authors propose an Attribute-Based Encryption naming scheme to tackle the management of content attributes in a distributed manner through an ontology-based management system and to enforce the access rules on public/cache-able routers through a set of name attributes. 
The main drawback of this scheme is that it is feasible only with flat naming schemes and cannot be used with hierarchical naming schemes.
Similarly, Grewe \textit{et al.}~\cite{grewe2017encircle} introduce an attribute-based encryption access control scheme for vehicular networks. The authors define two main components: (a) an access control management component to manage different policies and access rules based on their attributes, and (b) a cryptographic component to encrypt content based on the policies defined by the access rules. However, the placement of these components (centralized or distributed) and how they communicate/exchange attributes and policies remain an open question.
%

Feng \textit{et al.}~\cite{feng2018new} present a decentralized ciphertext-policy, attribute-based encryption mechanism to solve the revocation problem and preserve the consumers' privacy. In this work, the publisher creates an access control policy and encrypts the data according to this policy, current time, and the time that the data will be cached by routers. Then, it sends this policy in ciphertext in the network and hides the attribute values. Whereas, consumers can decrypt the ciphertext only if their attributes match the access structure in the ciphertext. 
The main drawback of this scheme is that the publisher needs to be always available.
Similary, Wu \textit{et al.}~\cite{wu2019chtds} propose a ciphertext policy, attribute-based encryption access control mechanism. In this scheme, the publisher splits the data into two sections: (i) the Public Content Section (PCS) that is the content itself, and (ii) the Content Key Section (CKS) that is a part of the key and specifies the data consumers. Consumers can decrypt the content by combining the PCS and the CKS. In doing so, they need to issue two interest packets, the first packet to request the PCS from the publisher or intermediate routers' content stores, while the second one to seek the CKS from the publisher. The main drawback of this work is that it requires the publisher to be constantly available which might not always be the case.
Another proxy-assisted access control scheme is proposed by Wu \textit{et al.}~\cite{wu2020multi} to ensure forward and backward security and achieve the user and attribute revocation. This mechanism includes an attribute authorization centre that is capable of creating an attribute of public keys. The publisher generates the access control policy for attributes and encrypts the content which is disseminated in the network and cached by NDN routers. Consumers can get the encrypted content from routers or publisher by inquiring keys from the central authorization center, and get the partially decrypted content from NDN routers that is decrypted using the proxy key. Consumers can then decrypt the remaining part to obtain the plaintext data. The main drawback of this work is that the increase in the number of attributes highly increases the proxy decryption time and ciphertext re-encryption.

Ramani \textit{et al.}~\cite{ramani2019ndn} introduce an alternative NDN signature design using attribute-based signatures to retrieve certificates and preserve the anonymity of individual publishers. In this approach, the publisher generates a policy by combining all attributes, then it creates a signature and adds it to the data packet. The consumer can verify the data based on two pieces of information: the data itself and the public key. The main limitation of this approach is that the verification process results in high cost, while the number of attributes increases the signature size.
In \cite{borgh2017employing}, Borgh \textit{et al.} extend Attribute-Based Encryption (ABE) for an IoT-based ICN architecture. Authors discussed two potential solutions. The first one is an ABE solution performed through a central authority. Sensors in this design encrypt the content through symmetric keys (encrypted by the central authority), which is then published through the ICN network. Users can access the content by decrypting both the symmetric keys and the content. The second solution is an ABE sensor system. In this solution, the central authority is not needed for encryption, since sensors perform ABE operations. 
The main drawback of these solutions is that they both use small-sized policies since sensors have limited computing resources. 


\subsection{Name-based Access Control}
A Name-based Access Control (NAC) mechanism~\cite{zhang2017nac} provides an automated key management process and content confidentiality using an encryption algorithm based only on the content name.
Hamdane \textit{et al.}~\cite{hamdane2013access} propose an enhancement-based access control scheme by eliminating the use of access control lists and uses a new cryptographic model. The content is encrypted by namespace keys, which are available to all nodes under the same namespace. However, the proposed scheme did not account for dynamic changes and large scale networks.
Wang \textit{et al.}~\cite{wang2014session} designed a session-based access control solution for ICN. This solution uses two different names for each piece of content, public and secure names. Secure names can be identified only only by authorized users. The authors demonstrate their solution through the example of an Online Social Network (OSN), where each user communicates with OSN over a session, and he/she is associated with a unique key shared with the service. The content uploading process requires the user to be authorized by the OSN using the shared key. Therefore, a user receives content over a session as well as the symmetric key and the required metadata to decrypt the content. All this information is encrypted by the session key. The main drawback of this mechanism is the overhead associated with several replicas of content that may be available across the network.

Hamdane \textit{et al.}~\cite{hamdane2013data} propose a data-based access control mechanism based on using encryption and lock password. In this approach, the access rights are established by the namespaces concept and specified in the access control list. This approach also provides a revocation mechanism by creating and encrypting a random key with public keys. The access control list is updated, and the revocation time is saved with the old password by the trusted server. However, creating and generating all the networks to add new access rights may become a challenging task.

\subsection{Identity-based Access Control}
In identity-based access control, the content consumer may prove its identity before requesting/consuming the content. The content publisher in return authenticates the requester before providing the content. Public or derivable keys are used to identify the consumers at the publisher level.
Wood \textit{et al.}~\cite{wood2014flexible} introduce a secure content distribution architecture on top of content-centric networks. This architecture is based on an identity-based cryptography and proxy re-encryption to provide simple public keys. These keys can be associated with users and pre-defined access rules. 
Although this mechanism may provide secure end-to-end communication, it requires the producer to be always connected, otherwise, consumers will not be able to decrypt the content.
Fotiou \textit{et al.}~\cite{fotiou2018rendezvous} design an access control mechanism for publish-subscribe based on enforcing the access control policies by Rendezvous Points (RVs). A content owner creates an access control policy that is assigned and protected with an ID. This ID is used as a public key to protect RVs. The owner encrypts the content through a symmetric encryption key, which is then encrypted and decrypted through the IBE encryption algorithm and the ID. Hence, the owner stores encrypted contents, the encrypted keys, the decrypted keys, and the list of authorized subscriber identities. When a publisher receives content, it advertises content information to RVs. If a subscriber needs to get the content, it sends a subscription message to the RV to initiates an authentication process. If the subscriber identity is included in the list of authorized identities, the RV sends a notification to the publisher in order to deliver the content to the subscriber. The main drawback of this solution is that it incurs compute-intensive operations due to the IBE algorithm, which cannot be used in constrained devices such as IoT.
Hamdane \textit{et al.}~\cite{hamdane2015credential} proposed a credential-based access control scheme for NDN. The proposed scheme assigns access rights to consumers based on certified credentials provided by an Access Control Manager (ACM), which is an entity that handles the private key(s) associated with the network namespace and defines the access and management rules. The communication and all security related operations are based on this centralized entity raising scalability concerns in the case of large-scale networks. Moreover, the proposed system requires an ACM for each namespace, the communication between ACMs in the case of different namespaces is an open question.
Similarly, Tseng \textit{et al.}~\cite{tseng2018fgac} introduce a fine-grained access control mechanism for NDN based on four roles: Key Generation Center, Content Producer, Online Shop, and Customers. The Key Generation Center is responsible for generating secret keys for the users and the Content Producer is able to disseminate the encrypted content in the network. The Online Shop allows customers to access the content. It is worth mentioning that the selection of trusted Online Shops maybe a challenging task.

\subsection{Re-encryption-based Access Control}
A re-encryption scheme~\cite{son2014conditional} may use two or more encryption operations to provide authentication or update the access rules~\cite{son2017new}.
Mangili \textit{et al.}~\cite{mangili2015cache} design an encryption-based extension for ICN. The proposed scheme aims to: (a) enforce confidential data dissemination: the producer encrypts the content, while the intermediate nodes cache encrypted rather than plaintext content; (b) track content access: consumers are authenticated by the original producer and fetch the required decryption keys; and (c) support policy evolution: producers may update the access policy through key-derivation and re-encryption. Although the proposed scheme is able to update the access policies after publishing the content, there is no guarantee to update all cached content instances. Also, the system is based on the availability of the original producer to retrieve keys for content decryption. 
Zheng \textit{et al.}~\cite{zheng2015achieving} introduced a dual-phase encryption mechanism that combines an one-time decryption key, proxy re-encryption, and all-or-nothing transformation. The original producer encrypts the original content with a key derived from its private key. When a consumer sends a request for specific content, the associated edge router re-encrypts the content (which is already encrypted by the producer) with a random key (each content has a different random key). The consumer is required to use both keys from the original producer and edge router to decrypt the content. Hence, the producer may control who can access the content. This scheme requires two keys for encryption/decryption, this the scalability is an open question. Moreover, the system is based on the availability of the original producer and the trustability of edge routers.


\subsection{Broadcast-based Access Control}
A broadcast-based access control tends to provide distributed access control to a group of users. The publisher encrypts the content and broadcasts the access rules to the network where only legitimate consumers may decrypt it.
Misra \textit{et al.}~\cite{misra2013secure} introduce a broadcast-based access control solution to secure content delivery in ICN. The content is encrypted by the producer using a public key, then the producer broadcasts this key into the network. The client can decrypt the content only if its symmetric key is matched with the public key. The main drawback of this scheme is that the number of keys broadcast in the network may be large severely affecting the network bandwidth.
Misra \textit{et al.}~\cite{misra2017accconf} tackled the limitation of requiring access control rules to be retrieved only from the provider. Hence, they propose an access control framework that allows a legitimate user to access and consume content directly from in-network caches without the need to be authenticated directly by the producer. The authors designed a broadcast encryption mechanism; the original producer broadcasts content to a set of legitimate consumers in a secure manner. However, a change in legitimate consumers (e.g, revoking the access privileges of a consumer, adding new consumers) is an open issue and requires further investigation.

\subsubsection{Probabilistic-based Access Control}
A probabilistic-based access control scheme integrates a probability function to determine if a request is authorized to access the content. 
Chen \textit{et al.}~\cite{chen2014encryption} proposed a probabilistic encryption-based access control mechanism for video streaming services over NDN. The proposed mechanism supports symmetric/asymmetric cryptographic operations to encrypt video content. The authors used Bloom filters to store the public keys of authorized consumers in order to filter invalid requests. The main drawbacks of this mechanism are the need for an always online producer and the impact of Bloom filter false positive errors. 

\subsection{Insights \& Learned Lessons}
One of the promising features of NDN is decoupling the content from its production location, which leads to in-network caching. The use of encryption-based access control mechanisms requires that all access rules need to be executed/enforced at the original producer level. This type of solution necessitates the original producer to remain connected and reachable at all times.

Some of the key lessons learned that should be noted are the following:
\begin{itemize}
	\item Dynamic content is widely used on the Internet. Access control schemes should be able to generate policies and rules upon the creation of the content with minimal overhead.
	\item Decoupling access control rules from content may help to identify unauthorized consumers, however, this solution violates ICN primitives, where the security-related information should travel with the content and be cached with it.
\end{itemize}

\section{Encryption Independent Access Control}
\label{sec:enc-ind}
In this section, we review existing encryption independent access control solutions. We refer to an access control solution as encryption independent if the access rules are determined independently from any underlying encryption. Existing solutions have been further classified into subcategories based on how rules are defined. Table~\ref{tab:encryption_independent_ac} summarizes the reviewed solutions.

\begin{table}[!t]
	\centering
	\caption{Summary of Encryption Independent access control mechanisms.}
	\label{tab:encryption_independent_ac}
	\tiny
	\begin{tabular}{cp{0.29\textwidth}
			cp{0.07\textwidth}cccp{0.28\textwidth}}
		\toprule
		
		\textbf{Ref.} &
		\makecell[c]{\textbf{Approach}} &
		\textbf{Level} &
		\textbf{Content} &
		\textbf{Cache} &
		\textbf{O/H} &
		\textbf{Comp.} &
		\makecell[c]{\textbf{Major Drawbacks}} \\
		\midrule
		
		\multicolumn{8}{c}{\textit{\textbf{Broker-based Access Control}}} \\ \midrule
		
		\cite{fotiou2012access} &
		$\bullet$ Use of AC Provider to coordinate access rules & 
		N &
		Static &
		\xmark &
		H &
		H &
		$\bullet$ System security relies on the AC Provider \newline
		$\bullet$ Requires extra overhead that may affect network delay \\
		
		\cite{singh2012trust}&
		$\bullet$ Use of broker to store different access policies & 
		N &
		Static &
		\xmark &
		H &
		H &
		$\bullet$ System security relies on the broker \newline
		$\bullet$ Requires identification/verification of subscribers/publishers \\
		
		\cite{zhu2020t} &
		$\bullet$ Lightweight time-based access control. & 
		N &
		Static &
		\xmark &
		H &
		H &
		$\bullet$ Scalability issues with mobility  \\ 

		\cite{xue2019secure} &
		$\bullet$ Edge-based access control framework & 
		N &
		Static &
		\cmark &
		L &
		L &
		$\bullet$ Performance based on edge node \\ \midrule
		
		\multicolumn{8}{c}{\textit{\textbf{In-network Cache Enforcement Access Control}}} \\ \midrule
		
		\cite{tan2014copyright}&
		$\bullet$ Enforces copyright when retrieving cached content & 
		P, R &
		Static &
		\xmark &
		H &
		H &
		$\bullet$ Caching of split content is not feasible in real scenarios \newline
		$\bullet$ Necessitates the availability of original producer \\
		
		\cite{kurihara2016consumer}&
		$\bullet$ Bypass censorship using consumer-driven access control & 
		P, R &
		Static &
		\xmark &
		H &
		H &
		$\bullet$ Solution is not feasible in off-path mode \\
		
		\cite{marxer2016access} &
		$\bullet$ Content-attendant policies in named function networks & 
		C, P &
		Static &
		\xmark &
		H &
		H &
		$\bullet$ Extra communication overhead to identify consumers \newline
		$\bullet$ Necessitates the availability of original producer \\
		
		\cite{liu2019cdac} &
		$\bullet$ Collaborative data access control scheme & 
		C, P &
		Static &
		\cmark &
		H &
		H &
		$\bullet$ Overhead the network \\ 

		\cite{lyu2020sbac} &
		$\bullet$ Blockchain matching-based access control & 
		P &
		Static &
		\cmark &
		H &
		H &
		$\bullet$ Requires publisher to be always online \\ \midrule
		
		\multicolumn{8}{c}{\textit{\textbf{Manifest-based Access Control}}} \\ \midrule
		
		\cite{kuriharay2015encryption}&
		$\bullet$ Decouples content and access rules using a manifest file & 
		P, R &
		Static &
		\xmark &
		H &
		M &
		$\bullet$ Access rules are not cached with content \newline
		$\bullet$ Necessitates the availability of original producer \\
		\midrule
		
		\multicolumn{8}{c}{\textit{\textbf{Interest-based Access Control}}} \\ \midrule
		
		\cite{ghali2015interest}&
		$\bullet$ Use of interest packets to enforce access rules & 
		C, P &
		Static &
		\xmark &
		H &
		M &
		$\bullet$ Requires mutual trust between consumer and provider \newline
		$\bullet$ Identical content with multiple names remains an open issue \\
		\midrule
		
		\multicolumn{8}{c}{\textit{\textbf{Signature-based Access Control}}} \\ \midrule
		
		\cite{li2015live}&
		$\bullet$ Lightweight integrity verification & 
		P, R &
		Static &
		\cmark &
		M &
		L &
		$\bullet$ Token refresh mechanism is required \\
		
		\bottomrule
	\end{tabular}
	\caption*{\it \scriptsize * N: Network C: Consumer, P: Producer, R: Router | L: Low, M: Medium, H: High}
\end{table}

\subsection{Broker-based Access Control}
In broker-based access control, a third party entity is used as a broker for the communication to apply and verify access control rules. The security of the whole system is based on the trust of this third party entity.
Fotiou \textit{et al.}~\cite{fotiou2012access} introduced an access control mechanism for ICN to protect the information of consumers and preserve their privacy. The proposed mechanism secures the information of consumers based on access control policies that are created by an Access Control Provider (ACP), who interacts with publishers, Rendezvous Nodes (RNs), and subscribers. Each publisher sends its access control policy to the ACP to have a URI assigned to it and then forwards the content to the RN. The RN sends the policy URI to the subscribers that request the content and, at the same time, the RN sends the URI of the associated policy to the ACP. Subsequently, the ACP verifies the policy and notifies the RN if the subscriber is allowed to access the content. If the verification is successful, the RN forwards the content to the subscriber. This mechanism produces additional computation and communication overhead at RNs that may increase the response latency. It also raises scalability concerns as the scale of the network grows.
Singh \textit{et al.}~\cite{singh2012trust} presented an access control approach for pub/sub networks. This approach secures the information based on access control policies specified by a local broker, who is an entity that verifies consumers and producers, as well as manages the constraints and the associated access levels. Both consumers and producers need to register with their local broker and define their credentials and attributes before starting the communication. This approach results in considerable overhead, while the access level identification/verification and the creation and management of the network between producers and the broker was not discussed.

Zhu \textit{et al.}~\cite{zhu2020t} design a lightweight, time-based content access control mechanism based on the notion of content subscription times. This approach is based on three cryptographic techniques: proxy re-encryption, identity-based encryption, and broadcast method, which allow the provider to encrypt their content, and push it to distribution servers. Then distribution servers re-encrypt, broadcast, and distribute this content to the network. In this approach, consumers can decrypt the content based on the private key and the specified subscription-time. However, this approach did not consider mobility and scalability.
On the other hand, Xue \textit{et al.}~\cite{xue2019secure} propose an edge-based access control framework for ICN to push the access control at the network edge and block unauthorized requests. The authors also propose a lightweight, privacy-preserving authentication protocol for the communication between consumers and edge routers based on group signature and hash chain techniques. This approach provides a revocation method, but results in considerable overhead without considering access control among different providers.

\subsection{Cache Enforcement-based Access Control}
In-network caching is a building blocks of NDN that improves the overall network performance. Both content caching and cache hits occur in a transparent way without informing the original producer or requester. The producer loses control over who can cache the content, which results in privacy, ownership, and copyright issues. Cache enforcement access control schemes aim to address this issue.
%
%
Tan \textit{et al.}~\cite{tan2014copyright} tackled the uncontrolled in-network caching problem for copyright enforcement during content retrieval from content stores. Authors argue that solutions based on encryption cannot protect content copyrights during caching. Hence, they proposed splitting large sized content into two parts: a big part and a small part with the constraint that without the small part no one can rebuild the whole content. The authors suggest that any node in the network can cache the big part of the content, while the small part remains at the original producer. To rebuild the content, the authors use bit-wise OR operations. The main drawback of this scheme is that the producer must be continuously connected to control the access policies.
%
Kurihara \textit{et al.}~\cite{kurihara2016consumer} designed an anonymization system to bypass censorship in content-centric networks. The authors designed a consumer-driven access control scheme that incorporates encryption-based access control into interest names and allows the producer to recycle specific content cached at intermediate content stores along the communication path. Plain-text routable names are used in the initial phase of communication, while encrypted names are used to provide anonymous communication. The main drawback of the proposed system is that it targets only on-path intermediate nodes, while cached content off the communication path cannot be recycled.

Marxer \textit{et al.}~\cite{marxer2016access} presented a set of content-attendant policies to complement content-based security principles in Named Function Networking (NFN) computations. In this approach, each content piece has an Access Control List (ACL) that contains the list of consumer identities that are permitted to access the content. Consumers send interests that have their public keys (identities) attached. When a provider receives such interests and determines that a consumer's public key matches a key in the ACL, the provider sends back the requested content (symmetrically encrypted) along with a symmetric key (asymmetrically encrypted with the consumer's public key). When the consumer receives the symmetric key and the content, it first decrypts the symmetric key and then uses this key to decrypt and access the content. A consumer can also request and retrieve the ACL of a content piece.
In this way, the consumer can become an independent content provider after receiving the content, ACL, and symmetric key. This approach, however, did not address the exchange of encrypted information in highly dynamic networks, such as vehicular networks. 

Liu \textit{et al.}~\cite{liu2019cdac} design a collaborative data access control scheme for NDN, where access control function is carried out at routers that may cache content instead of a single content producer. In this approach, the producer encrypts the content using a symmetric key. It then disseminates the key along with a unique sub-key for this content to the routers. When a consumer needs to access the content, it sends an Interest packet to request the encrypted content from the producer or a router. After that, the consumer sends another Interest packet to the cache-enabled router requesting the sub-key in order to decrypt the content. However, the authors did not discuss the generation of sub-keys, while their solution requires additional Interest packets in order to retrieve the content and the encryption keys.
In another work, Lyu \textit{et al.}~\cite{lyu2020sbac} propose a matching-based access control model based on Blockchain to ensure the security of sharing, auditing, and revocation of the content's publisher. In this work, the publisher defines the permitting operations for each content, classifies the requesting nodes based on their attributes, and identifies the access right of the requesting nodes. This access right is transferred in the form of transactions on the Blockchain to ensure the security of transmission and accurate record keeping of access activities. This approach also allows routers to cache content in ICN only if the share operation is included in the permitting operations, and any requester can decrypt the shareable content by using the transaction of access taken for shareable content. However, the publisher needs to be always available to provide the operations for requesters and content.

\subsection{Manifest-based Access Control}
Manifest-based Access Control aims at providing a separate file (manifest) that specifies the access policy and rules, thus decoupling the access control rules from the original content. The main objective of this type of access control is to ensure minimum communication overhead and maximum utilization of in-network caches.
Kuriharay \textit{et al.}~\cite{kuriharay2015encryption} proposed an encryption-based access control mechanism for content-centric networks. This scheme is designed from the perspective of manifest-based control retrieval. The idea consists of securing the content manifests and decoupling the encrypted content from access rules. To retrieve a content piece, a consumer needs to get authorized by the original content producer and download an encrypted content copy. Then, the consumer will use the manifest file to locate the required keys to decrypt the content. The main drawback of this scheme is that both the original producer and a consumer need to be simultaneously connected. This contrasts the design principles of ICN/NDN, where communication shall be session-less, the content is decoupled from the location it was originally produced, and the content consumption may be asynchronous compared to its production.


\subsection{Interest-based Access Control}
Interest-based Access Control mechanisms consist of attaching information to interest packets, which is used to determine the access rules for the requested content. Therefore, access control can be enforced even for cached content.
Ghali \textit{et al.}~\cite{ghali2015interest} designed an interest-based access control scheme to enforce access rules through the use of information in interest packets. In this scheme, access control rules and content-encryption are decoupled; the original content producer has the ability to enforce any access control rules without dealing with content encryption or key distribution. This scheme also supports both hash- and encryption-based name obfuscation, so that sensitive content names cannot be predicted by unauthorized entities. Moreover, mutual trust verification between routers and consumers was proposed to authorize access to locally cached content and overcome interest replay attacks. Despite the advantages of using obfuscated content names, the same content may be cached multiple times under different names, which can negatively impact the distribution of cache resources.

\subsection{Signature-based Access Control}
Signature-based Access Control mechanisms consist of a universal content signature verification process to enforce access control rules.
Li \textit{et al.}~\cite{li2015live} extended the NDN architecture and proposed a lightweight integrity verification architecture. The proposed architecture uses one-way hash functions based on the Merkle Hash Tree algorithm to produce content signatures and generate tokens to sign and verify content pieces. The proposed algorithms allow the original content producer to control the content access rules by distributing tokens to authorized consumers. However, this approach requires efficient mechanisms for token renewal and group key management.

\subsection{Insights \& Learned Lessons}
Determining access control rules for specific content independently of underlying encryption algorithms overcomes the requirement of having producers always online, so that authorized access to content can be achieved even if producers have been disconnected. The network becomes aware of the AC rules and policies as well as updates the rules for already cached contents.

The key lessons learned are the following:
\begin{itemize}
	\item The use of broker-based mechanisms can help to enforce access rules and content security, however, the trustability of the system depends on trusting the broker node.
	\item Original producers may be able to update access rules even if the content is cached across the network. To this end, cache enforcement schemes are desirable to evict content associated with outdated access rules and cache content with up-to-date access rules.
	\item Re-encryption and broadcast-based schemes can increase the security and trust levels. However, they result in significant computation overheads and consume large amounts of in-network cache resources.
\end{itemize}

\section{Challenges \& Future Research Directions}
\label{sec:directions}
The main objective of this work is to collect, categorize, and analyze different access control mechanisms in NDN/ICN as well as to identify key challenges and provide potential research directions based on the presented analysis and lessons learned. In this section, we elaborate on different challenges and provide insights and possible research directions on core aspects.

\subsection{Encryption-Based Access Control}
Attribute-based access control mechanisms result in fine-grained policies, offering access control enforcement to a group of users. This granularity may come at the cost of having a centralized authority, probably affecting the scalability of the system. Approaches to enhance the scalability of attribute-based access control are definitely worth exploring. Name-based access control mechanisms provide semantically meaningful access control that matches the name-based nature of NDN, while identity-based access control mechanisms result in clean designs which, however, may require compute-intensive cryptographic operations. Re-encryption-based access control offers the capability to update access control rules, however, it may result in limited scalability, since it requires two or more cryptographic operations for providing authentication and updates of the rules (as already used in cloud environments \cite{Son2014}). Finally, broadcast-based access control provides distributed access to a group of users; however, the resulting overhead may be substantial, showcasing the need for approaches (e.g., scoped broadcast) that could mitigate the overhead. In a nutshell, further investigation is needed to optimize encryption-based access control for NDN-driven application paradigms.

\subsection{Encryption-Independent Access Control}
Broker-based access control results in simple designs; however, the security of the whole system relies on trusting and securing the broker. Mechanisms to identify security compromises of brokers and to establish trust with them are highly valuable. Cache enforcement access control mechanisms may contribute to enforcing the ownership and copyright rules when it comes to content cached in the network, while manifest-based access control can reduce the communication overhead and maximize the utilization of in-network caching. Similarly, interest-based access control may obfuscate content names; however, this may result in having the same content cached in the network under multiple names. This issue can be mitigated through intelligent cache management strategies which could identify multiple copies of the same content. Finally, signature-based access control may allow producers to control the content access rules by assigning access tokens to consumers. However, managing these tokens may be cumbersome, a direction that requires further research.

\subsection{Dynamic Attribute-based Access Control}
An attribute-based access control mechanism performs well in an ICN-based network. However, such a mechanism requires certain features to be adopted in real-world scenarios and large-scale networks. Generating dynamic attributes based on the content and type of communication is required, along with providing distributed computation rather than a proxy or a third-party entity. A possible solution might be the use of machine learning to generate attributes and Blockchain to decentralize the computation and processing~\cite{jan2020security}.

\subsection{Publisher-Subscriber Communication}
Most of today's applications are based on a subscription model. After the subscribers register for a specific topic, generated content will be retrieved by them in an autonomous manner. Ensuring that only the authorized subscribers receive the content is vital. 
Evicted subscribers should not be able to decrypt content after the eviction. At the same time, when the content may not use the provider's public key for encryption purposes, dynamically generated content based on active subscriptions may be a solution. Logical key hierarchy mechanisms can be used as an alternative solution to generate keys based on the dynamic changes of the subscription group.

\subsection{Resource-Constrained Devices}
IoT is considered to be the future of the Internet, where things communicate with/over the Internet using any network, any service, from any place. Most IoT sensors have limitations in computation and memory~\cite{rehman2020ccic, mastorakis2020dlwiot}. To this end, lightweight access control and cryptographic mechanisms are needed for IoT devices. Directions of interest may include symmetric encryption/decryption schemes as well as evaluations of the trade-offs between security guarantees and cost. Furthermore, compute-intensive cryptographic operations can be outsourced from the resource-constrained IoT devices to powerful nodes. 

\subsection{In-network Cache Update Enforcement}
Although in-network caching aims at enhancing network performance, decoupling the content from its original location leads to losing control of the content after its publication. The producer should be able to control and manage who is authorized to access the content even if the content is already cached in the network. Enforcing such policies is much needed in today's Internet. The use of Blockchain and other collaborative approaches may help to achieve this goal; however, a flexible trust model is required with minimal computation and overhead. Approaches based on network controllers that keep track of cached content and enforce the selected policies to the network may also be explored.

\section{Conclusion}
\label{sec:conclusion}

NDN follows session-less communication and implements content-based security where security is applied directly to the content itself rather than the communication channel.
Access Control is a fundamental security requirement aiming to enforce access to content only by authorized consumers. Content naming and the decoupling of content from its original locations lead to different issues in realizing access control in NDN compared to solutions for the traditional TCP/IP network architecture \cite{nour2019coexistence}.

In this survey, we presented and discussed a variety of access control mechanisms in NDN. We first gave an overview of the ICN paradigm, its characteristics, features, and then introduced the NDN architecture. Subsequently, we described the paradigm shift from session-based security to content-based security along with different cryptographic algorithms, security protocols, and introduction to access control. We also presented existing NDN-based access control mechanisms and classified them into different categories. Finally, we presented and identified research gaps and challenges that may be considered by the research community when designing efficient access control solutions. It is important to note that the reviewed access control mechanisms in NDN need to be seen in different contexts depending on the types of applications they are applied to. For instance, the broker-based access control mechanisms can be effective but could incur considerable overhead on the network. Similarly, the cryptographic access control (including signature-based access control) techniques are promising for non-realtime NDN applications. Further research on access control mechanisms should be conducted, especially, when we consider resource-constrained devices and publisher-subscriber communication. We hope that our work will act as a cornerstone for further research in developing efficient, viable, and effective access control mechanisms in NDN, as well as help the community to understand the trade-offs and merit of existing access control solutions.

\bibliographystyle{ACM-Reference-Format}
\bibliography{Ref}

\end{document}